\documentclass[twocolumn,showpacs,preprintnumbers,amsmath,amssymb,pra]{revtex4}
\usepackage{graphicx}
\begin{document}

\title{Inferring periodic orbits from spectra of simple shaped micro-lasers}

\author{M. Lebental$^{1,2}$, N.
  Djellali$^{1}$, C. Arnaud$^{1}$, J.-S. Lauret$^{1}$, J. Zyss$^1$\\
R. Dubertrand$^2$, \fbox{C. Schmit$^2$}, and E. Bogomolny$^2$}
\email{lebental@lpqm.ens-cachan.fr} \affiliation{$^1$CNRS, Ecole
Normale Sup\'erieure de Cachan,  UMR 8537,
Laboratoire de Photonique Quantique et Mol\'eculaire, 94235 Cachan, France \\
$^2$ CNRS, Universit\'e Paris Sud,  UMR 8626,\\ Laboratoire de
Physique Th\'eorique et Mod\`eles Statistiques, 91405 Orsay, France}

\date{\today}

\begin{abstract}
Dielectric micro-cavities are widely used as laser resonators and
characterizations of their spectra are of interest for various
applications. We experimentally investigate micro-lasers of simple
shapes (Fabry-Perot, square, pentagon, and disk). Their lasing
spectra consist mainly of almost equidistant peaks and the distance
between peaks reveals the length of a quantized periodic orbit. To
measure this length with a good precision, it is necessary to take
into account different sources of refractive index dispersion.  Our
experimental and numerical results agree with the superscar model
describing the formation of long-lived states in polygonal cavities.
The limitations of the two-dimensional approximation are briefly
discussed in connection with micro-disks.
\end{abstract}


\maketitle

\section{Introduction}

Two-dimensional micro-resonators and micro-lasers are being
developed as building blocks for optical telecommunications
\cite{siegman, vahalalivre}. Furthermore they are of interest as
sensors for chemical or biological applications \cite{vahalalivre,
driessencapteur, eaulourde} as well as billiard toy models for
quantum chaos \cite{stonescience, nouspra}. Towards fundamental and
applied considerations, their spectrum is one of the main features.
It was used, for instance, to experimentally recover some
information about the refractive index \cite{vardenydisque} or
geometrical parameters \cite{vahalaanneau}.

In this paper we focus on cavities much larger than the wavelength
and propose to account for spectra in terms of periodic orbit
families. Cavities of the simplest and most currently used shapes
were investigated: the Fabry-Perot resonator, polygonal cavities
such as square and pentagon, and circular cavities.

\begin{figure}[ht!]
\begin{minipage}[t!]{.33\linewidth}
\includegraphics[width=1\linewidth]{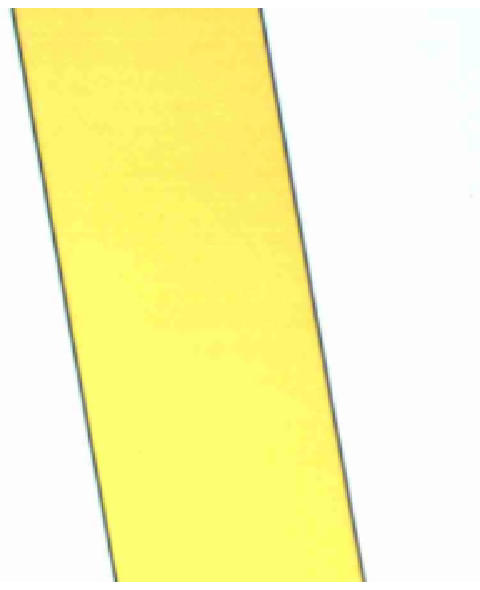}
\end{minipage}\hfill
\begin{minipage}[t!]{.33\linewidth}
\includegraphics[width=1\linewidth]{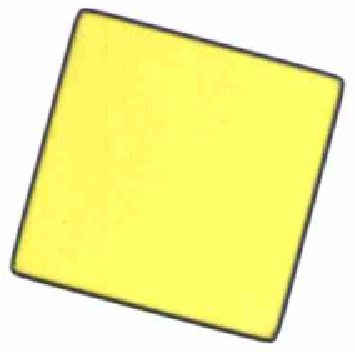}
\end{minipage}\hfill
\begin{minipage}[t!]{.33\linewidth}
\includegraphics[width=1\linewidth]{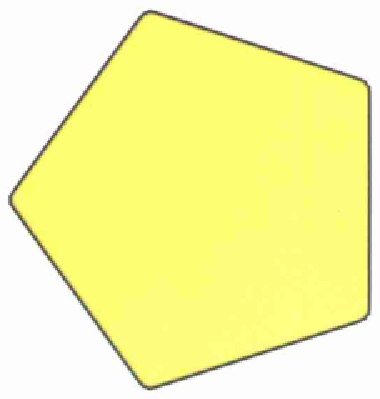}
\end{minipage}\hfill
\begin{minipage}[t!]{.33\linewidth}
\includegraphics[width=1\linewidth]{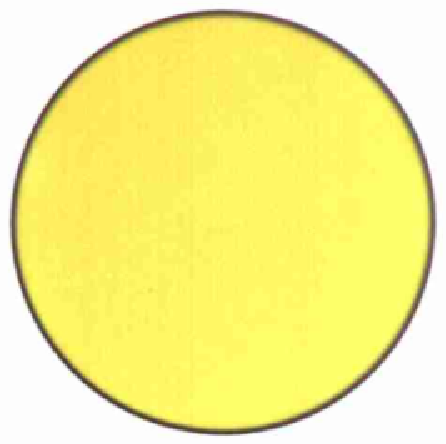}
\end{minipage}\hfill
\begin{minipage}[t!]{.33\linewidth}
\includegraphics[width=1\linewidth]{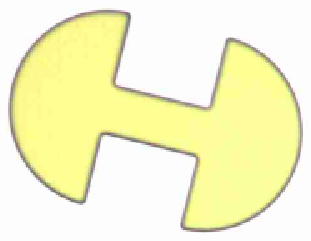}
\end{minipage}\hfill
\begin{minipage}[t!]{.33\linewidth}
\includegraphics[width=1\linewidth]{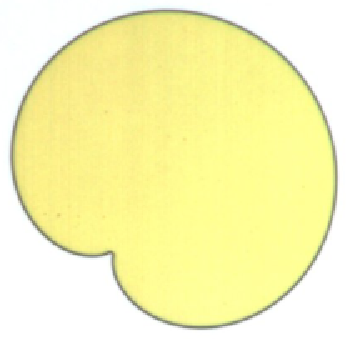}
\end{minipage}\hfill
\caption{Optical microscope photographs of some organic
micro-lasers: stripe (partial view, used as Fabry-Perot resonator),
    square, pentagon, disk, quasi-stadium, and cardioid. Typical dimension: $100~\mu
    m$.}
    \label{photos}
\end{figure}

Our experiments are based on quasi two-dimensional organic
micro-lasers \cite{APL}. The relatively straightforward fabrication
process ensures good quality and reproducibility as well as
versatility in shapes and sizes (see Fig.~\ref{photos}). The
experimental and theoretical approaches developed in this paper can
be easily extended to more complicated boundary shapes. Moreover
this method is useful towards other kinds of micro-resonators, as it
depends only on cavity shape and refractive index.

The paper is organized as follows. In Section~\ref{model} a
description of the two-dimensional model is provided together with
its advantages and limitations. In Section~\ref{fabry_perot}
micro-lasers in the form of a long stripe are investigated as
Fabry-Perot resonators to test the method and evaluate its
experimental precision.  This protocol is then further applied to
polygonal cavities. In Section~\ref{square_cavity} the case of
square cavities is discussed whereas in
Section~\ref{pentagonal_cavity} dielectric pentagonal cavities are
investigated. The theoretical predictions based on a superscar model
are compared to experiments as well as numerical simulations and a
good agreement is found. Finally, in Section~\ref{circular_cavity}
the case of several coexisting orbits is briefly dealt with on the
example of circular cavities.

\section{Preliminaries}\label{model}

Dielectric micro-cavities are quasi two-dimensional  objects whose
thickness is of the order of the wavelength but with much bigger
plane dimensions (see Fig.~\ref{photos}).  Although such cavities
have been investigated for a long time both with and without lasing,
their theoretical description is not quite satisfactory. In
particular, the authors are not aware of true three-dimensional
studies of high-excited electromagnetic fields even for passive
cavities. Usually one uses a two-dimensional approximation but its
validity is not under control.

\begin{figure}[ht!]
\begin{minipage}[t!]{.4\linewidth}
\includegraphics[width=0.99\linewidth]{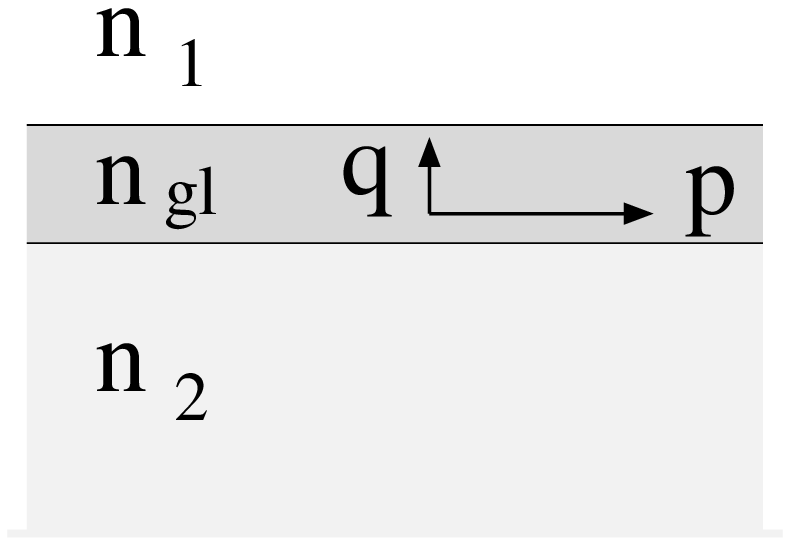}
\begin{center}(a)\end{center}
\end{minipage}\hfill
\begin{minipage}[t!]{.59\linewidth}
\includegraphics[width=0.99\linewidth]{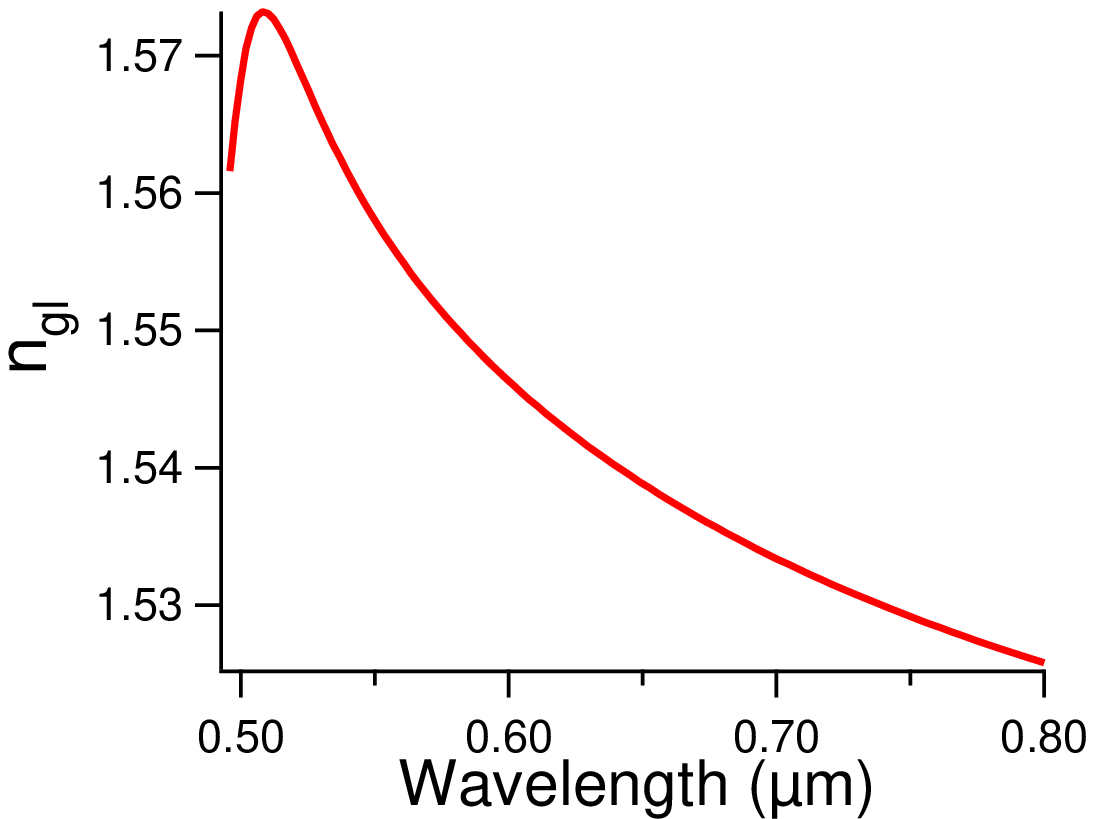}
\begin{center}(b)\end{center}
\end{minipage}
\caption{(a) Notations for refractive indexes and propagation
    wavenumbers. From top to bottom, the layers of our samples
    \cite{APL}
    are the air ($n_1=1$), a polymer (PMMA) doped with a laser dye
    (DCM) ($n_{gl}= 1.54$), and another polymer (SOG) ($n_2=1.42$)
    or silica ($n_2=1.45$). (b) Refractive index of the gain layer versus the wavelength
    inferred from ellipsometric measurements.}
    \label{notationsindices}
\end{figure}

Within such approximation fields inside the cavity and close to its
two-dimensional boundary are treated differently. In the bulk, one
considers electromagnetic fields as propagating inside an infinite
dielectric slab (gain layer) with refractive index $n_{gl}$
surrounded  by medias with refractive indices $n_1$ and $n_2$
smaller than $n_{gl}$. In our experiments, the gain layer is made of
a polymer (PMMA) doped with a laser dye (DCM) and sandwiched between
the air and a polymer (SOG) layer (see Fig.~\ref{notationsindices}
(a) and \cite{APL}). It is well known (see e.g. \cite{cavity} or
\cite{indiceeffectif}) that in such geometry there exist a finite
number of propagating modes confined inside the slab by total
internal reflection. The allowed values of transverse momentum
inside the slab, $q$, are determined from the standard relation
\begin{equation}
e^{2ihq}r_1r_2=1 \label{quantification}
\end{equation}
where $h$ is the slab thickness and $r_{1,2}$ are the Fresnel
reflection coefficients on the two horizontal interfaces. For total
internal reflection
\begin{equation}
r_i=\exp(-2{\rm i} \delta_i)
\label{reflection}
\end{equation}
where
\begin{equation}
\delta_i=\arctan\left(\nu_i\frac{\sqrt{n_{gl}^2\sin^2
      \theta-n_i^2}}{n_{gl}\cos \theta}\right) \label{phase}
\end{equation}
Here $\theta$ is the angle between the direction of wave propagation
inside the slab and the normal to the interface. The $\nu_{i}$
parameter is 1 (resp. $(n_{gl}/n_{i})^2$) when the magnetic field
(resp. the electric field) is perpendicular to the slab plane. The
first and second cases correspond respectively   to TE and TM
polarizations.

\begin{figure}
\begin{center}
\includegraphics[width=0.99\linewidth]{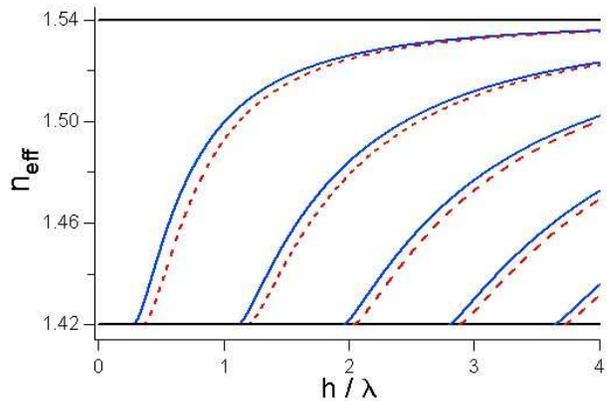}
\end{center}
\caption{ Effective refractive index
    versus the thickness over wavelength variable calculated from Eq. (\ref{effective}).
    The refractive indices are assumed
    to be constant: 1 for air, 1.42 for SOG, and 1.54 for the gain layer
    (horizontal black lines). The TE polarization is plotted with solid blue
    lines and TM polarization with dotted red lines. Integer  $l$ (see (\ref{effective}))
     increases from left to right starting from $l=0$.}
    \label{indice}
\end{figure}

Denoting the longitudinal momentum, $p= n_{gl}k\sin
\theta $, as $p=n_{eff}k$, the effective refractive index, $n_{eff}$,
is determined from the following dispersion relation
\begin{eqnarray}
&&2\pi \frac{h}{\lambda}\sqrt{n_{gl}^{2}-n_{eff}^{2}}= \arctan \left
  (\nu_1\frac{\sqrt{n_{eff}^{2}-n_{1}^{2}}}{\sqrt{n_{gl}^{2}-n_{eff}^{2}}}
\right )\nonumber\\ &+&\arctan \left
  (\nu_2\frac{\sqrt{n_{eff}^{2}-n_{2}^{2}}}{\sqrt{n_{gl}^{2}-n_{eff}^{2}}}
\right ) +l\pi\;,\;\;l\in \mathbb{N}\;. \label{effective}
\end{eqnarray}
This equation has only a finite number of propagating solutions
which can easily be obtained numerically. Fig.~\ref{indice} presents
possible propagating modes for our experimental setting $n_1=1$
(air), $n_2=1.42$ (SOG) \footnote{For some samples, the underlying
layer is silica with refractive index $n_2=1.45$, so $n_{eff}$ is
slightly different.} and $n_{gl}=1.54$ deduced from ellipsometric
measurements (see Fig.~\ref{notationsindices} (b)) in the
observation range.

The Maxwell equations for waves propagating inside the slab are thus reduced to the
two-dimensional scalar Helmholtz equation:
\begin{equation}
\left (\Delta+n_{eff}^2k^2 \right )\Psi_{in}(x,y)=0\;.
\label{helmholtz}
\end{equation}
$\Psi$ represents the field perpendicular to the slab, i.e. the
electric field for TM and the magnetic field for TE polarization
\footnote{This definition is consistent all over the paper. In the
literature, these names are sometimes permutated.}.

This equation adequately describes the wave propagation inside the
cavity. But when one of these propagating modes hits the cavity
boundary, it can partially escape from the cavity and partially be
reflected inside it. To describe correctly different components of
electromagnetic fields near the boundary, the full solution of the
three dimensional vectorial Maxwell equations is required, which to
the authors knowledge has not yet been adressed in this context.
Even the much simpler case of scalar scattering by a half-plane
plate with a small but finite thickness is reduced only to numerical
solution of the Wiener-Hopf type equation \cite{wiener}.

To avoid these complications, one usually considers that the fields
can be separated into TE and TM polarization and obey the scalar
Helmholtz equations (\ref{helmholtz})
\begin{equation}
\left (\Delta+n_{in,out}^2k^2 \right )\Psi_{in,out}(x,y)=0\;.
\label{Eqs}
\end{equation}
with $n_{in}$ is the $n_{eff}$ effective index inferred from Eq.
(\ref{effective}) and $n_{out}$ the refractive index of the
surrounding media, usually air so $n_{out}=1$. This system of
two-dimensional equations is closed by imposing the following
boundary conditions
\begin{equation}
\Psi_{in}|_{B}=\Psi_{out}|_B\;,\;\;\; \nu_{in}\frac{\partial
\Psi_{in}}{\partial \vec{\tau}}|_B= \nu_{out}\frac{\partial
\Psi_{out}}{\partial \vec{\tau}}|_B\;. \label{bc}
\end{equation}
Here $\vec{\tau}$ indicates the direction normal to the boundary and
$\nu$ depends on the polarization. When the electric (resp.
magnetic) field is perpendicular to the cavity plane, called TM
polarization (resp. TE polarization), $\nu_{in,out}=1$ (resp.
$\nu_{in,out}=1/n^2_{in,out}$). Notice that these definitions of
$\nu$ are not the same for horizontal and vertical interfaces.

We consider this standard two-dimensional approach keeping in mind
that waves propagating  close to the boundary (whispering gallery
modes) may deviate significantly from two-dimensional predictions.
In particular leakage through the third dimension could modify the
life-time estimation of quasi-stationary states.\\

Our polymer cavities are doped with a laser dye and uniformly pumped
one by one  from above \cite{APL}, so that the pumping process
induces no mode selection. The complete description of such lasing
cavities requires  the solution of the non-linear Maxwell-Bloch
equations (see e.g. \cite{harayama,tureci,tureci2}  and references
therein). For clarity, we accept here a  simplified point of view
(see e.g. \cite{siegman} Sect.~24) according to which true lasing
modes can be represented as a linear combination of the passive
modes which may lase (i.e. for which gain exceeds losses)
\begin{equation}
\Psi_{\rm lasing}=\sum_m C_m \Psi_m\;.
\end{equation}
From physical considerations, it is natural to consider the $\Psi_m$
modes as the quasi-stationary states of the passive cavity. Though
this choice leads to well known difficulties (see e.g.
\cite{siegman}) it is widely noticed and accepted at least for modes
with small losses (cf. \cite{tureci,harayama2,tureci2}).

For each individual lasing mode,  the $C_m$ coefficients could be
determined only after the solution of the full Maxwell-Bloch
equations. But due to the statistical nature of fluorescence the
lasing effect starts randomly and independently during each pump
pulse. So it is quite natural to average over many pump pulses. Then
the mean spectrum exhibits peaks at frequencies of all possible
lasing modes. The experimental data studied in this paper are
recorded after integration over 30 pump pulses and agree with this
simple statistical model. More refined verifications are in
progress.

\section{Fabry-Perot resonator}\label{fabry_perot}

The Fabry-Perot configuration is useful for the calibration control
of further spectral experiments due to the non ambiguous single
periodic orbit family which sustains the laser effect.

\begin{figure}[ht!]
\begin{minipage}[t!]{.5\linewidth}
\includegraphics[width=0.8\linewidth]{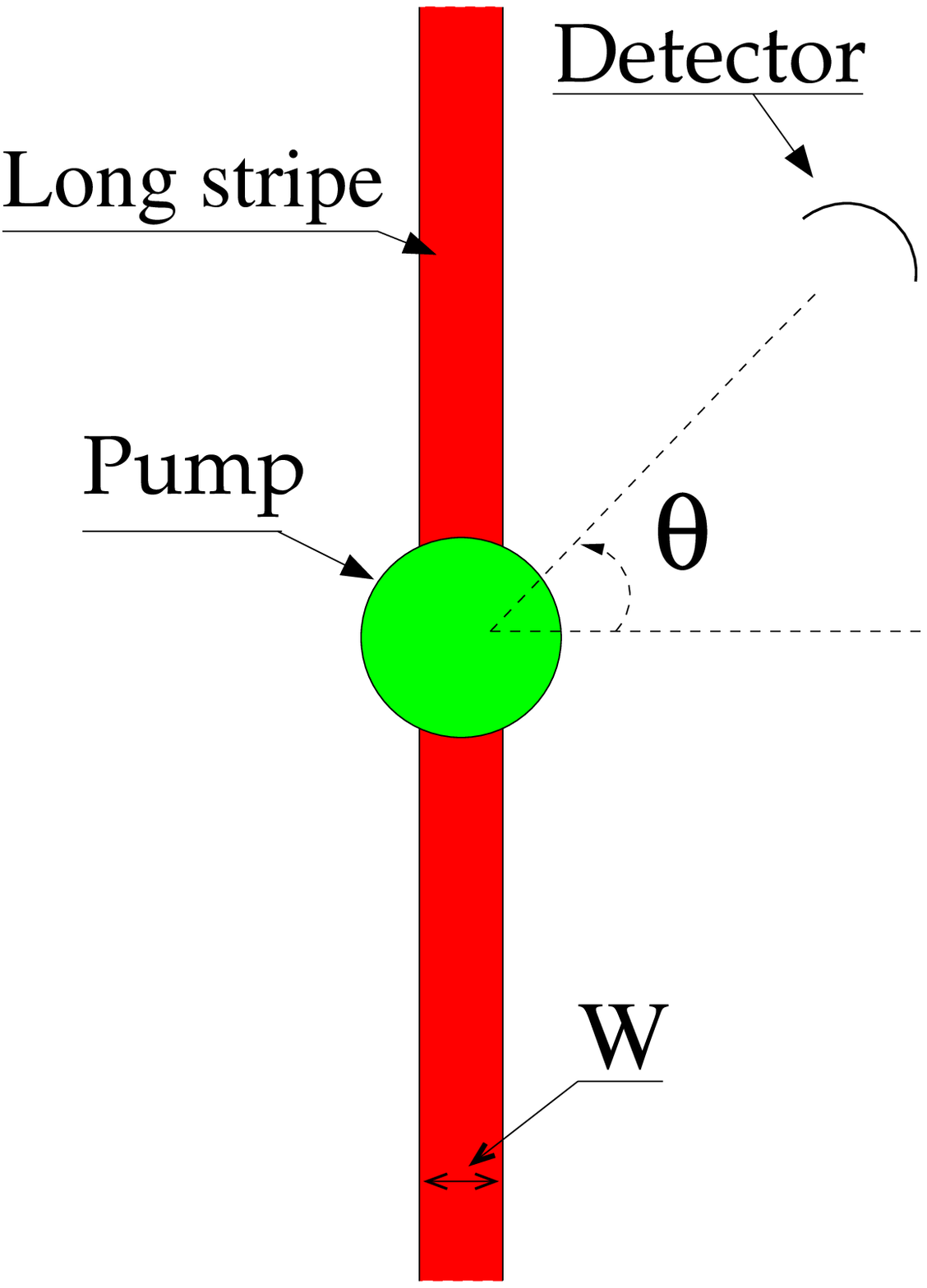}
\begin{center}(a)\end{center}
\end{minipage}\hfill
\begin{minipage}[t!]{.5\linewidth}
\includegraphics[width=0.99\linewidth]{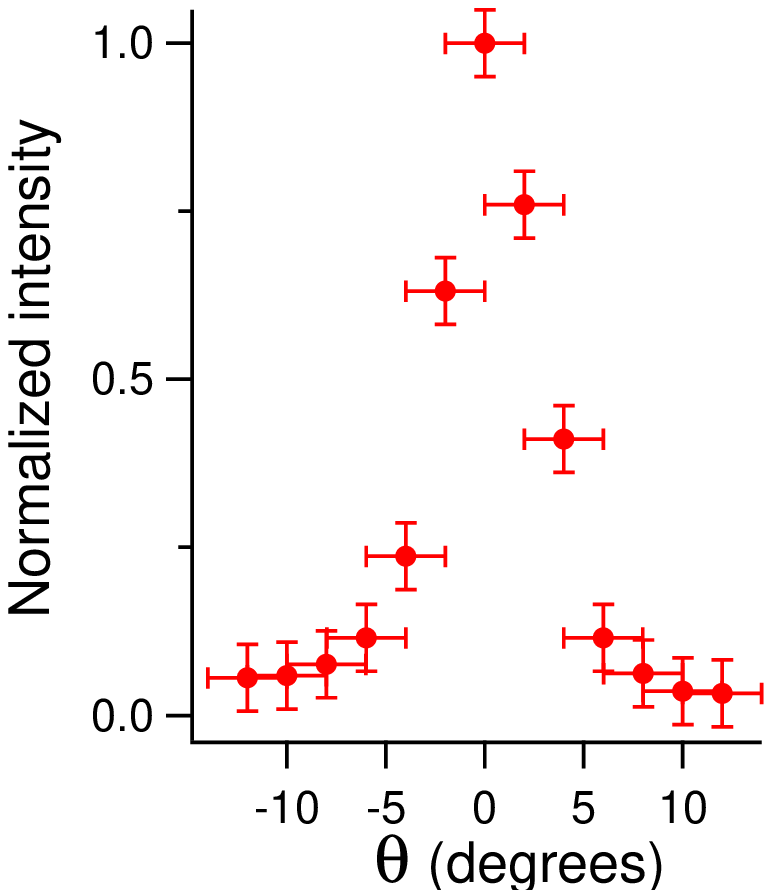}
\begin{center}(b)\end{center}
\end{minipage}
\caption{(a) Diagram summarizing the main features of the
Fabry-Perot experiment.
    (b) Detected intensity versus $\theta$ angle  for a Fabry-Perot experiment.}
    \label{fpschema}
\end{figure}

A long stripe can be considered to a good approximation as a
Fabry-Perot resonator. In fact the pumping area is very small
compared to the length (see Fig.~\ref{fpschema} (a)) and the
material is slightly absorbing, so that reflections at far
extremities can be neglected. Moreover the pumping area is larger
than the width of the stripe, thus the gain is uniformly distributed
over the section. For a Fabry-Perot cavity, the emission is expected
along both $\theta=0$ and $\theta=\pi$ directions (see
Fig.~\ref{fpschema} (a) for notations). Fig.~\ref{fpschema} (b)
shows that this directional emission is observed experimentally
which confirms the validity of our set-up.

\begin{figure}
\begin{minipage}[t!]{\linewidth}
\includegraphics[width=0.8\linewidth]{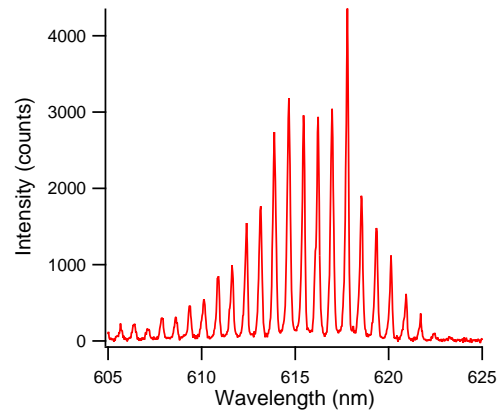}
\begin{center}(a)\end{center}
\end{minipage}\hfill
\begin{minipage}[t!]{\linewidth}
\includegraphics[width=0.8\linewidth]{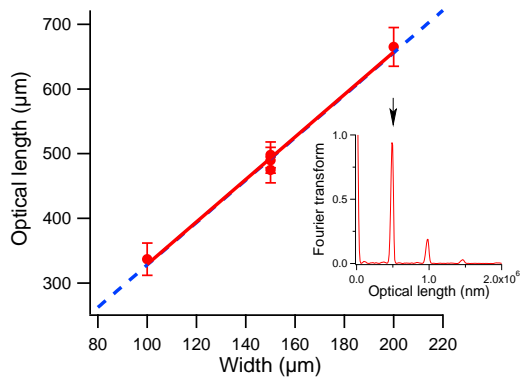}
\begin{center}(b)\end{center}
\end{minipage}
\caption{(a) Experimental spectrum of a Fabry-Perot resonator with
       $W=150~\mu m$.
       (b) Optical length versus Fabry-Perot width $W$. The experiments (red points) are
       linearly fitted by the solid red line. The dashed blue line
       corresponds to the theoretical prediction without any
       adjusted parameter.
       Inset: Normalized Fourier transform of the  spectrum in (a) expressed
       as intensity versus wavenumber.}
    \label{fpresultats}
\end{figure}

The experimental spectrum averaged over 30 pump pulses is made up of
almost regularly  spaced peaks (see Fig.~\ref{fpresultats} (a))
which is typically expected  for a Fabry-Perot resonator. In fact,
due to coherent effects, the $k$ wavenumbers of quasi-bound states
of a passive Fabry-Perot cavity are determined from the quantization
condition along the only periodic orbit of $L=2W$ length as for
(\ref{quantification}):
\begin{equation}
r^2{\rm e}^{{\rm i}~L~k~n_{eff}(k)}=1
\end{equation}
where $r$ is the Fresnel reflection coefficient and $n_{eff}$ is the
effective refractive index (\ref{effective}). The solutions of this
equation are complex numbers: the imaginary part corresponds to the
width of the resonance and the real part (called $k_m$ afterwards)
gives the position of a peak in the spectrum and verify
\begin{equation}
L~k_m~n_{eff}(k_m)= 2\pi~m\;,\hspace{0.5cm}m\in \mathbb{N}\;.
\end{equation}
With $\delta k_m=k_{m+1}-k_m$ assumed to be small, the distance between adjacent
peaks is constrained by
\begin{equation}
\delta k_m [n_{eff}(k_m)+k_m~\frac{\partial n_{eff}}{\partial
k}(k_m)]~L=2\pi\;.
\end{equation}
We call
\begin{equation}
n_{full}=n_{eff}(k_m)+k_m~\frac{\partial n_{eff}}{\partial k}(k_m)
\end{equation}
the full effective refractive index. It is a sum over two terms: one
corresponding to the phase velocity, $n_{eff}(k_m)$, and the other
one to the group velocity, $k_m~\frac{\partial n_{eff}}{\partial
k}(k_m)$. If $n_{full}$ is considered as a constant over the
observation range, which is true with a good accuracy, $\delta k$
can be retrieved from the experimental spectrum. For instance, the
Fourier transform of the spectrum (intensity versus $k$) is made up
of regularly spaced peaks (Fig.~\ref{fpresultats} (b) inset), with
the first one (indicated with an arrow) centered at the optical
length ($L~n_{full}$) and the others at its harmonics. So the
geometrical length of the periodic orbit can be experimentally
inferred from the knowledge of $n_{full}$ which is independently
determined as described below. For the Fabry-Perot resonator, the
geometrical length is known to be $2W$, thus allowing to check the
experimental precision. The relative statistical errors on the $W$
width is estimated to be less than 3 \%. The error bars in
Fig.~\ref{fpresultats} (b) are related to the first peak width of
the Fourier transform and are less than 5 \% of the optical length.

The  full effective refractive index, $n_{full}$, is independently
inferred from ellipsometric measurement (Fig.~\ref{notationsindices}
(b)) and standard effective index derivation described in the
previous Section. Depending on the parameter $h/\lambda$ (thickness
over wavelength), one or several modes are allowed to propagate. Our
samples are designed such as only one TE and TM modes exist with
$n_{eff}$ effective refractive index according to
Eq.~(\ref{effective}).

In Fig.~\ref{indice} the  refractive index of the gain layer,
$n_{gl}$, is assumed to be constant: $n_{gl}=1.54$ in the middle of
the experimental window, $\lambda$ varying from 0.58 to 0.65 $\mu$m.
From Eq.~(\ref{effective}) a $n_{eff}=1.50$ is obtained in the
observation range with a $h=0.6~\mu$m thickness, and corresponds to
the phase velocity term. The group velocity term $k_m~\frac{\partial
n_{eff}}{\partial k}(k_m)$ is made up of two dispersion
contributions: one from the effective index (about 4 \%) and the
other from the gain medium (about 7 \%). The dependance of $n_{gl}$
with the wavelength is determined with the GES 5 SOPRA ellipsometer
from a regression with the Winelli II software (correlation
coefficient: 0.9988) and plotted on Fig.~\ref{notationsindices} (b).
Taking into account all contributions (that means calculating the
effective refractive index with a dispersed $n_{gl}$), the
$n_{full}$ full effective refractive index is evaluated to be 1.645
$\pm$ 0.008 in the observation range. So the group velocity term
made up of the two types of refractive index dispersion contribute
for 10 \% to the full effective index, which is significant compared
to our experimental precision. The $n_{full}$ index depends only
smoothly on polarization (TE or TM), and on the $h$ thickness, which
is measured with a surface profilometer Veeco (Dektak$^3$ST). Thus,
the samples are designed with thickness $0.6~\mu$m and the precision
is reported on the full effective index which is assumed to be 1.64
with a relative precision of about 1 \% throughout this work.

Considering all of these parameters, we obtain a satisfactory
agreement between measured and calculated optical lengths, which
further improves when taking into account several Fabry-Perot
cavities with different widths as shown on Fig.~\ref{fpresultats}
(b). The excellent reproducibility (time to time and sample to
sample) is an additional confirmation of accuracy and validity. With
these Fabry-Perot resonators, we have demonstrated a spectral method
to recover the geometrical length of a periodic orbit which can now
be confidently applied to different shapes of micro-cavities.

\section{Square micro-cavity}\label{square_cavity}

\begin{figure}
\begin{minipage}[t!]{\linewidth}
\includegraphics[width=0.8\linewidth]{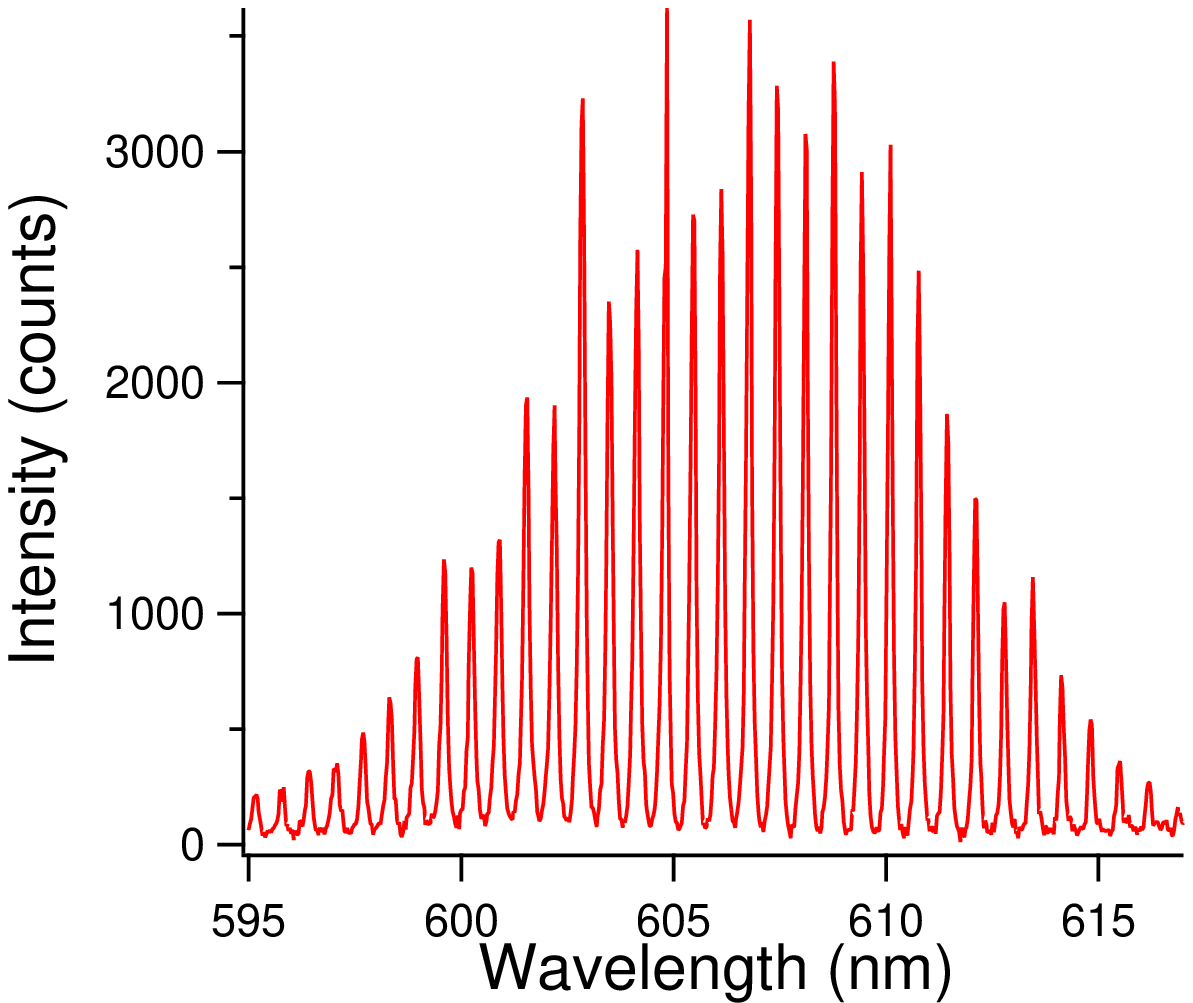}
\begin{center}(a)\end{center}
\end{minipage}\hfill
\begin{minipage}[t!]{\linewidth}
\includegraphics[width=0.8\linewidth]{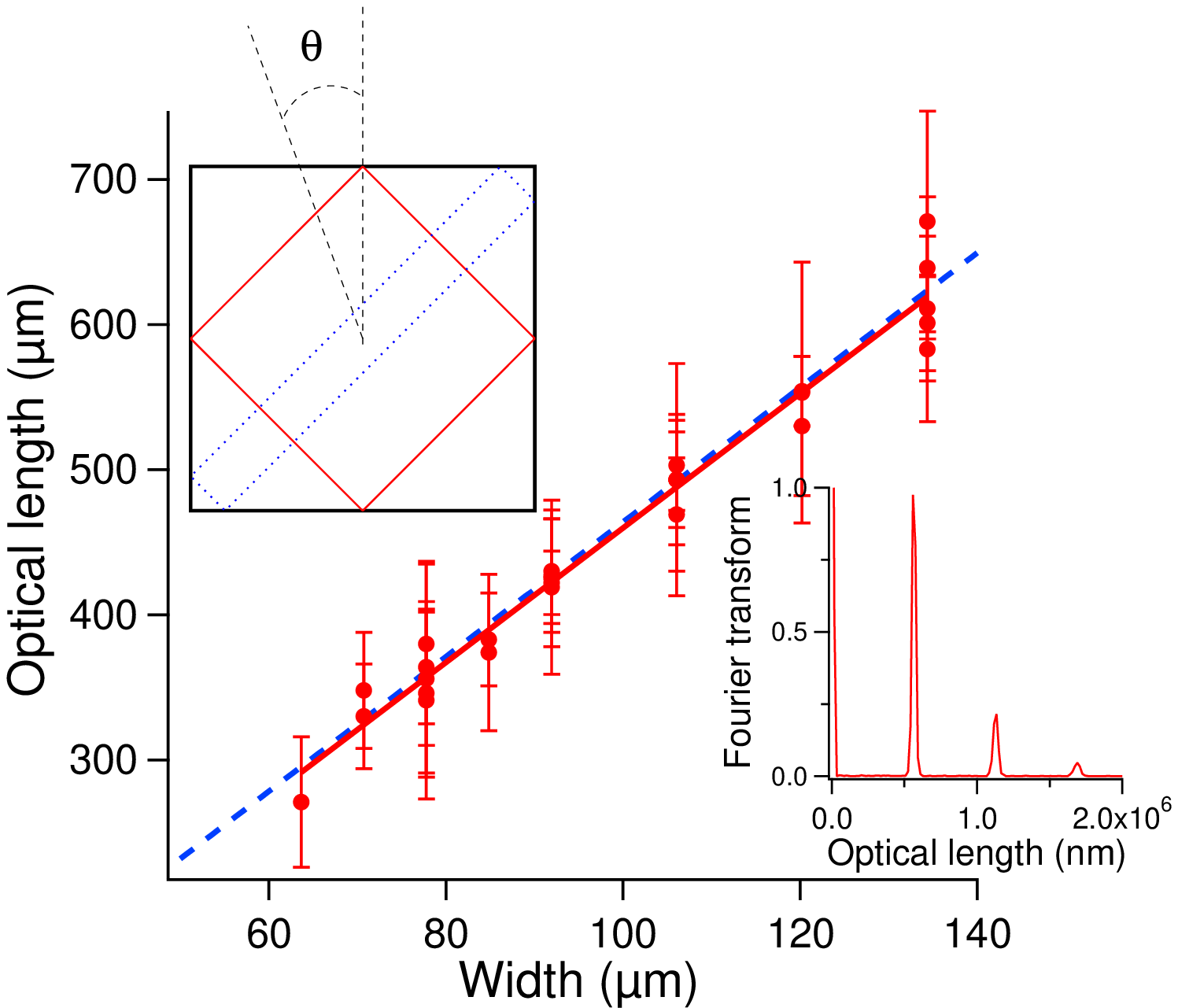}
\begin{center}(b)\end{center}
\end{minipage}
\caption{(a) Experimental spectrum of a square-shaped micro-laser of
$135~\mu m$ side width.
    (b) Optical length versus  $a$ square side width. The experiments (red points) are
       linearly fitted by the solid red line. The dashed blue line
       corresponds to the theoretical prediction (diamond periodic orbit) without any
       adjusted parameter.
      Top inset: Two representations of the diamond periodic orbit.
       Bottom inset: Normalized Fourier transform of the spectrum in (a) expressed as
       intensity versus wavenumber.}
    \label{carrespectre}
\end{figure}

In the context of this paper square-shaped micro-cavities present a
double advantage. Firstly, they are increasingly used in optical
telecommunications \cite{vahalalivre,pooncarre}. Secondly, the
precision and validity of the parameters used above can be tested
independently since there is only one totally confined periodic
orbit family. In fact the refractive index is quite low (about 1.5),
so the diamond (see Fig.~\ref{carrespectre} (b), top inset) is the
only short-period orbit without refraction loss (i.e. all reflection
angles at the boundary are larger than the critical angle
$\chi_c=\arcsin(1/n)\approx 42^{\circ}$.)

In a square-shaped cavity light escapes mainly at the corners due to
diffraction. Thus the quality design of corners is critical for the
directionality of emission but not for the spectrum. Indeed for
reasonably well designed squared micro-cavity (see
Fig.~\ref{photos}), no displacement of the spectrum peaks is
detectable by changing the $\theta$ observation angle. The spectra
used in this paper are thus recorded in the direction of maximal
intensity.

Fig.~\ref{carrespectre} (a) presents a typical spectrum of a
square-shaped micro-cavity. The peaks are narrower than in the
Fabry-Perot resonator spectrum, indicating a better confinement, as
well as regularly spaced, revealing a single periodic orbit. Data
processing is performed exactly as presented in the previous
Section: for each cavity the Fourier transform of the spectrum is
calculated (Fig.~\ref{carrespectre} (b), bottom inset) and the
position of its first peak is located at the optical length.
Fig.~\ref{carrespectre} (b) summarizes the results for about twenty
different micro-squares, namely: the optical length inferred from
the Fourier transform versus the $a$ square side width. These
experimental results are fitted by the solid red line. The dotted
blue line corresponds to an a priori slope given by $n_{full}$
(1.64) times the geometrical length of the diamond periodic orbit
($L=2\sqrt{2}a$). The excellent agreement confirms that the diamond
periodic orbit family provides a dominant contribution
to the quantization of dielectric square resonator.\\

This result is far from obvious as square dielectric cavities are
not integrable. At first glance the observed dominance of one
short-period orbit can be understood from general considerations
based on trace formulae which are a standard tool in semiclassical
quantization of closed multi-dimensional systems (see e.g.
\cite{LesHouches,prange} and references therein). In general trace
formulae express the density of states (and other quantities as
well) as a sum over classical periodic orbits. For two-dimensional
closed cavities
\begin{equation}
d(k)\equiv\sum_{n}\delta(k-k_n)\approx \sum_p c_p {\rm e}^{{\rm i}kL_p-{\rm
    i}\mu_p}+{\rm c.c. }
\label{tf}
\end{equation}
where $k$ is the wavenumber and $k_n$ are the eigenvalues of a
closed cavity. The summation on the right part is performed over all
periodic orbits labeled by $p$. $L_p$ is the length of the $p$
periodic orbit, $\mu_p$ is a certain phase accumulated from
reflection on boundaries and caustics, and amplitude $c_p$ can be
computed from classical mechanics. In general for integrable and
pseudo-integrable systems (e.g. polygonal billiards) classical
periodic orbits form continuous periodic orbit families and in two
dimensions
\begin{equation}
c_p\sim \frac{A_p}{\sqrt{L_p}}
\end{equation}
where $A_p$ is the geometrical area covered by a periodic orbit
family (see the example of circular cavities in Section
\ref{circular_cavity}).

Non-classical contributions from diffractive orbits  and different
types of creeping waves (in particular, lateral waves \cite{prange})
are individually smaller by a certain power of $1/k$ and are
negligible in semiclassical limit $k\to\infty$ compared to periodic
orbits.

There exist no true bound states for open systems. One can only
compute the spectrum of complex eigenfrequencies  of
quasi-stationary states. The real parts of such eigenvalues give the
positions of resonances and their imaginary part measure the losses
due to the leakage from the cavity.

For such systems it is quite natural to assume that the density of
quasi-stationary states
\begin{equation}
d(k)\equiv\frac{1}{\pi}\sum_{n}\frac{ {\rm Im}(k_{n})}{(k-{\rm
Re}(k_{n}))^{2} +{\rm Im}(k_{n})^{2}} \label{quasi_density}
\end{equation}
can be written in a form similar to (\ref{tf}) but the contribution
of each periodic orbit has to be multiplied by the product of all
reflection coefficients along this orbit (as it was done in a
slightly different problem in \cite{prange})
\begin{equation}
d(k)\approx \sum_p c_p \left [\prod_{j=1}^{N_p}r_p^{(j)}\right ]{\rm e}^{{\rm i}kL_p-{\rm
    i}\mu_p}+{\rm c.c. }\;.
\label{trace_k}
\end{equation}
Here $N_p$ is the number of reflections at the boundary and
$r_p^{(j)}$ is the value of  reflection coefficient corresponding to
the $j^{{\rm th}}$ reflection for the $p$ periodic orbit.

When the incident angle is larger than the critical angle the
modulus of the reflection coefficient equals 1 (see Eq.
(\ref{reflection})), but  if a periodic orbit hits a piece of
boundary with angle smaller than the critical angle, then $|r_p|<1$
thus reducing the contribution of this orbit. Therefore, the
dominant contribution to the trace formula for open dielectric
cavities is given by short-period orbits ($c_p\propto1/\sqrt{L_p} $)
which are confined by total internal reflection. For a square cavity
with $n=1.5$ the diamond orbit is the only confined short-period
orbit which explains our experimentally observation of its
dominance.

Nevertheless, this reasoning is incomplete because the summation of
contributions of one periodic orbit and its repetitions in polygonal cavities
does not produce a complex pole which is the characteristics of
quasi-stationary states.

\begin{figure}
\begin{minipage}[t!]{.99\linewidth}
\includegraphics[ angle=-90, width=0.8\linewidth]{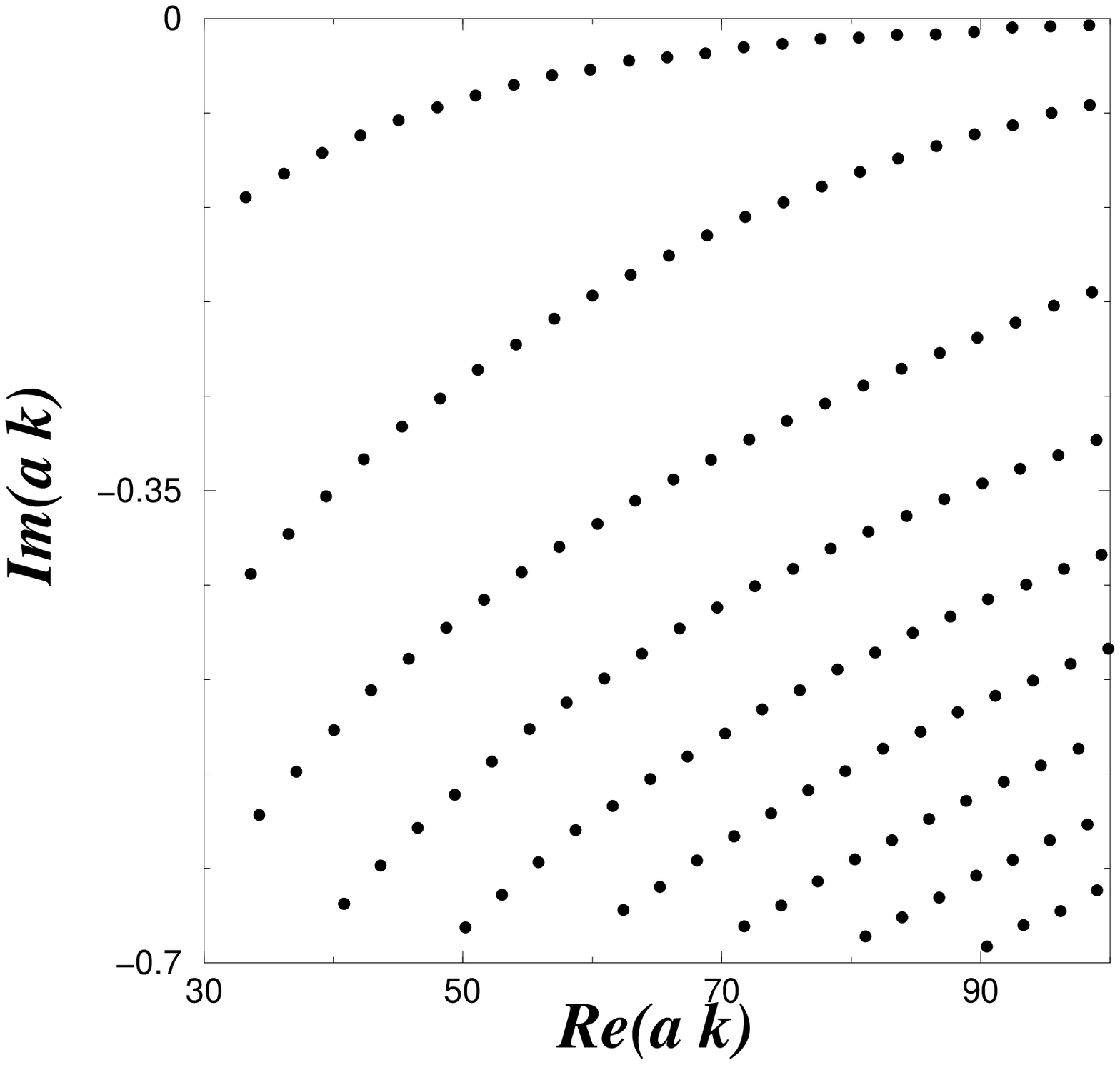}
\begin{center}(a)\end{center}
\end{minipage}\hfill
\begin{minipage}[t!]{.99\linewidth}
\includegraphics[width=0.8\linewidth]{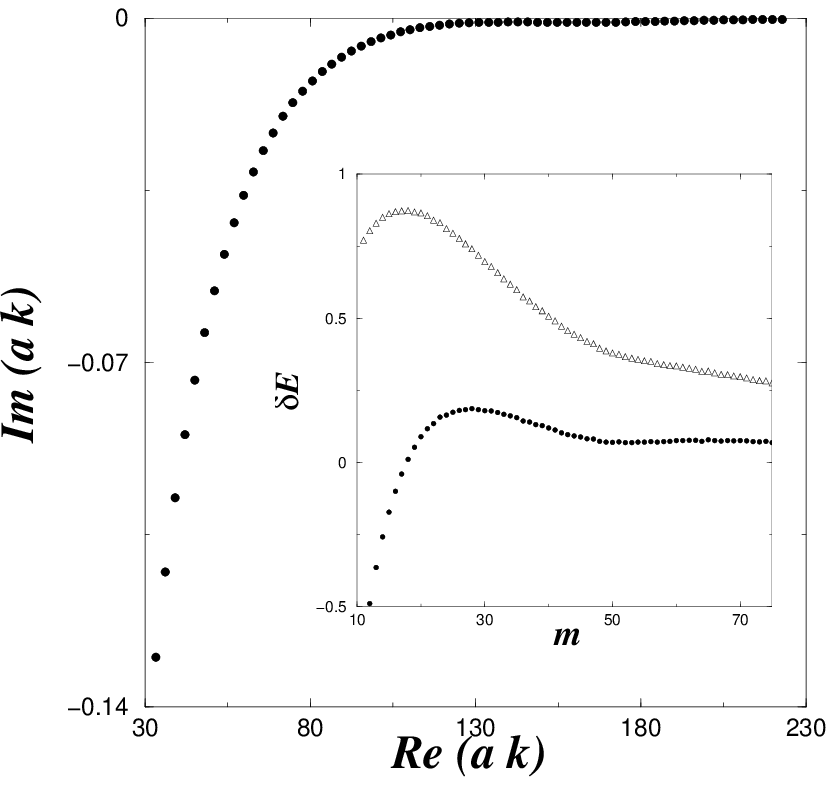}
\begin{center}(b)\end{center}
\end{minipage}
\caption{(a) Imaginary parts versus real parts of the wavenumbers
 of quasi-stationary states with $(-\; -)$ symmetry for a dielectric
square resonator with $n_{eff}=1.5$ surrounded by air with $n=1$.
(b) The same as in (a) but for the states with the smallest modulus
of the imaginary part (the most confined states). Inset. Empty
triangles: the difference (\ref{deltaE})  between the real part of
these wavenumbers and the asymptotic expression. Filled circles: the
same but when the correction term (\ref{correction}) is taken into
account.} \label{total}
\end{figure}

In order to better understand the situation, we have performed
numerical simulations for passive square cavities in a
two-dimensional approximation with TM polarization (see
Section~\ref{model} and \cite{wiersig}). Due to symmetries, the
quasi-stationary eigenstates can be classified according to
different parities with respect to the square diagonals. In
Fig.~\ref{total} (a), the imaginary parts of wavenumbers are plotted
versus their real part for states antisymmetric according to the
diagonals (that means obeying the Dirichlet boundary conditions
along the diagonals) and called here $(-\; -)$ states.

\begin{figure*}
\begin{minipage}{1.\linewidth}
\begin{minipage}[t!]{.33\linewidth}
\includegraphics[width=.99\linewidth]{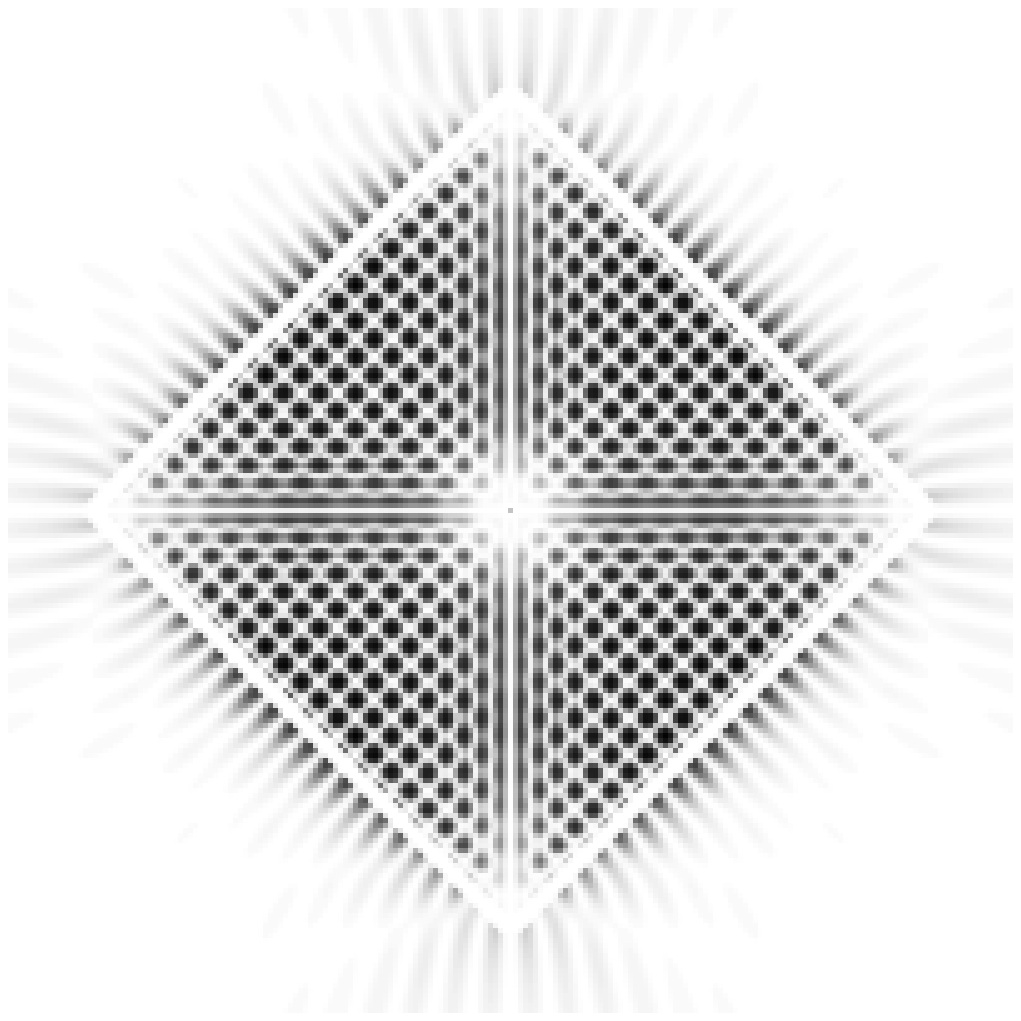}
\begin{center}(a)\end{center}
\end{minipage}\hfill
\begin{minipage}[t!]{.33\linewidth}
\includegraphics[width=.99\linewidth]{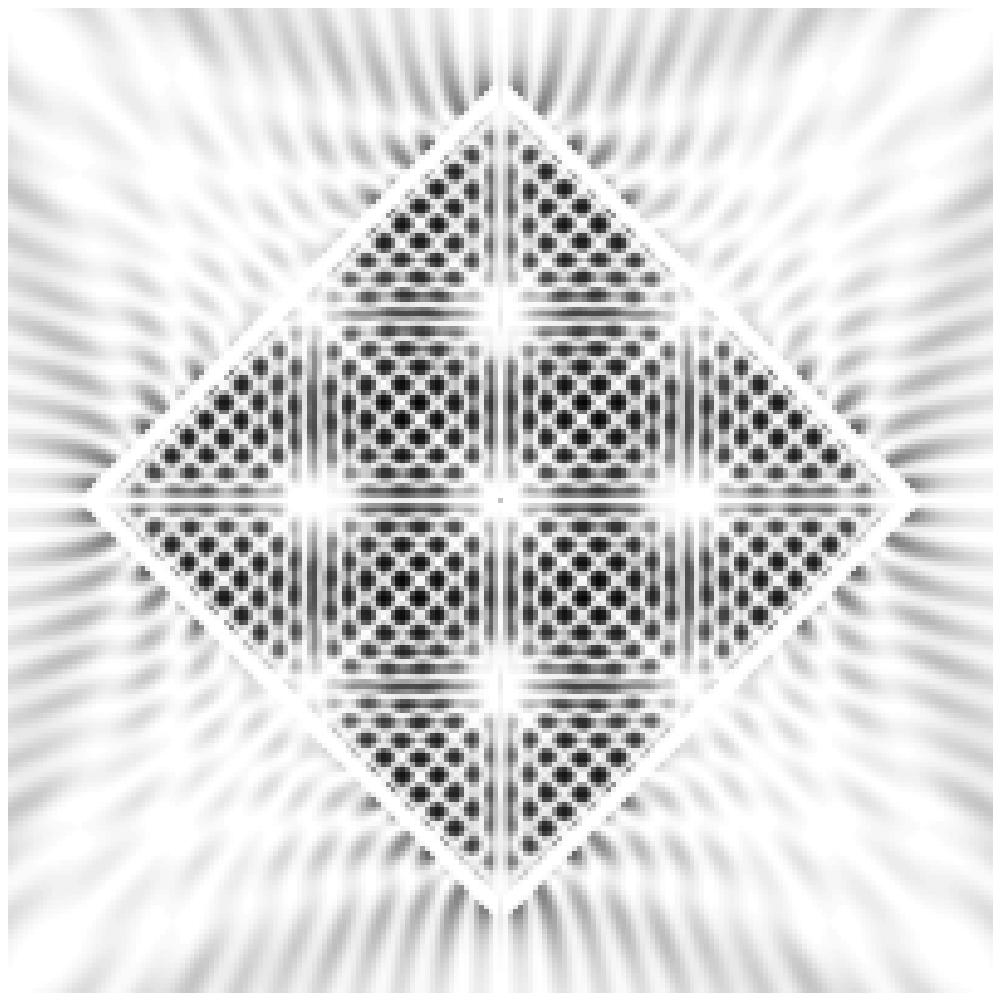}
\begin{center}(b)\end{center}
\end{minipage}\hfill
\begin{minipage}[t!]{.33\linewidth}
\includegraphics[width=.99\linewidth]{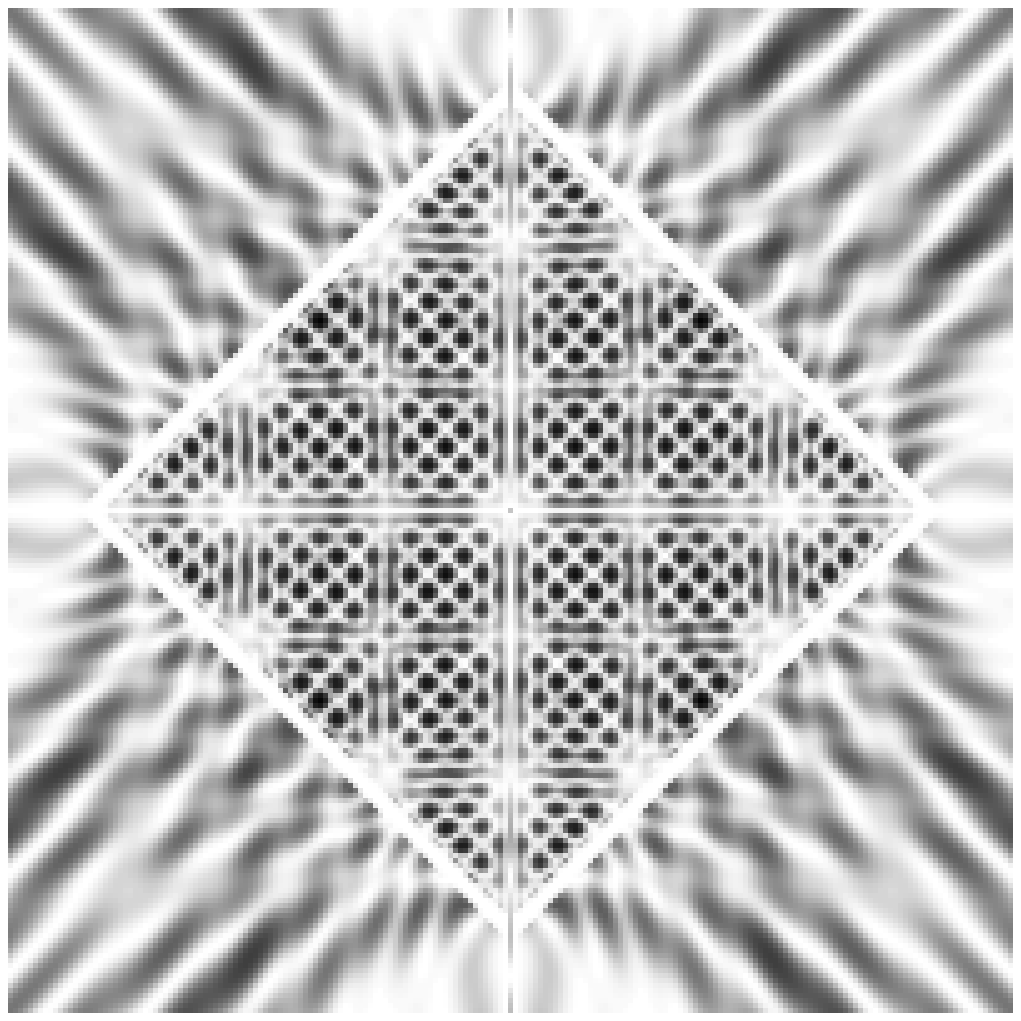}
\begin{center}(c)\end{center}
\end{minipage}
\caption{Squared modulus of wave functions with $-\; -$ symmetry
calculated with numerical simulations. (a)
  $ak=68.74-.026~{\rm i}$, (b)  $ak=68.84-.16~{\rm i}$, (c)
  $ak=69.18-.33~{\rm i}$.}
\label{psi}
\begin{minipage}[t!]{.33\linewidth}
\begin{center}
\includegraphics[width=.79\linewidth]{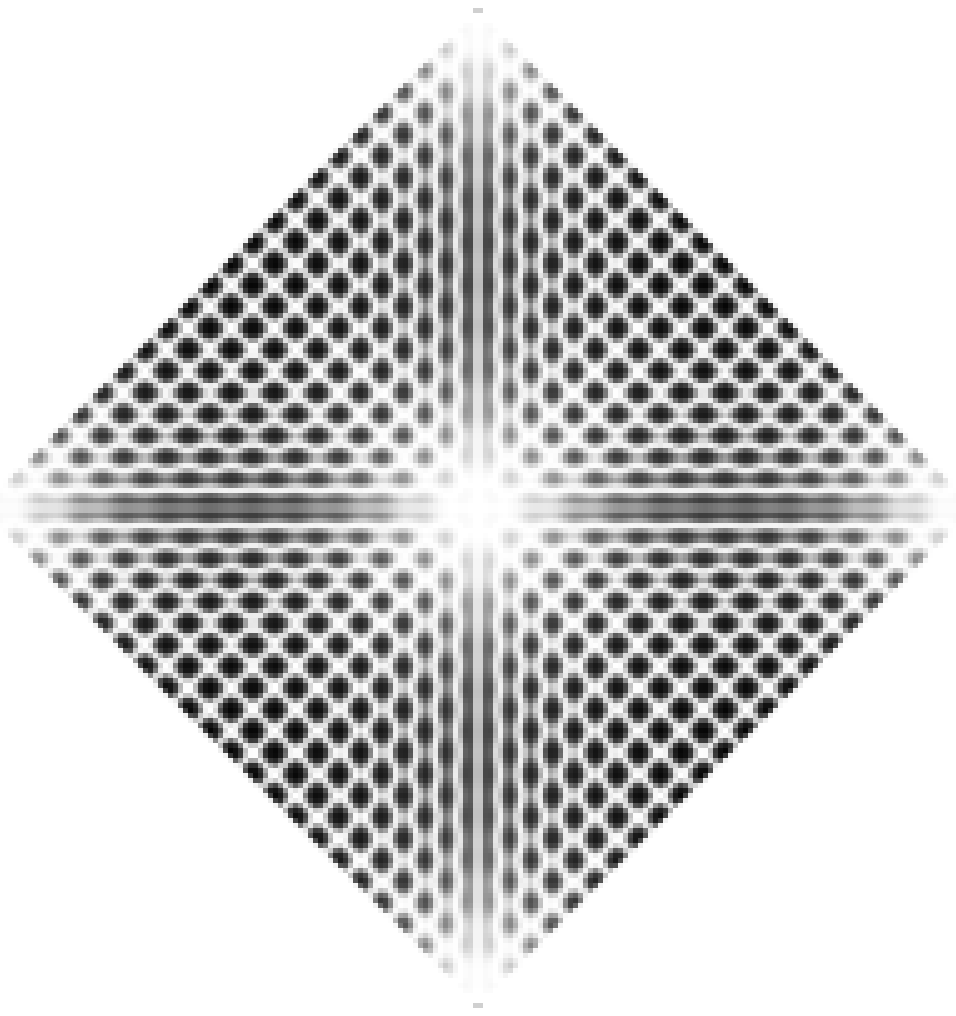}
\end{center}
\begin{center}(a)\end{center}
\end{minipage}\hfill
\begin{minipage}[t!]{.33\linewidth}
\begin{center}
\includegraphics[width=.79\linewidth]{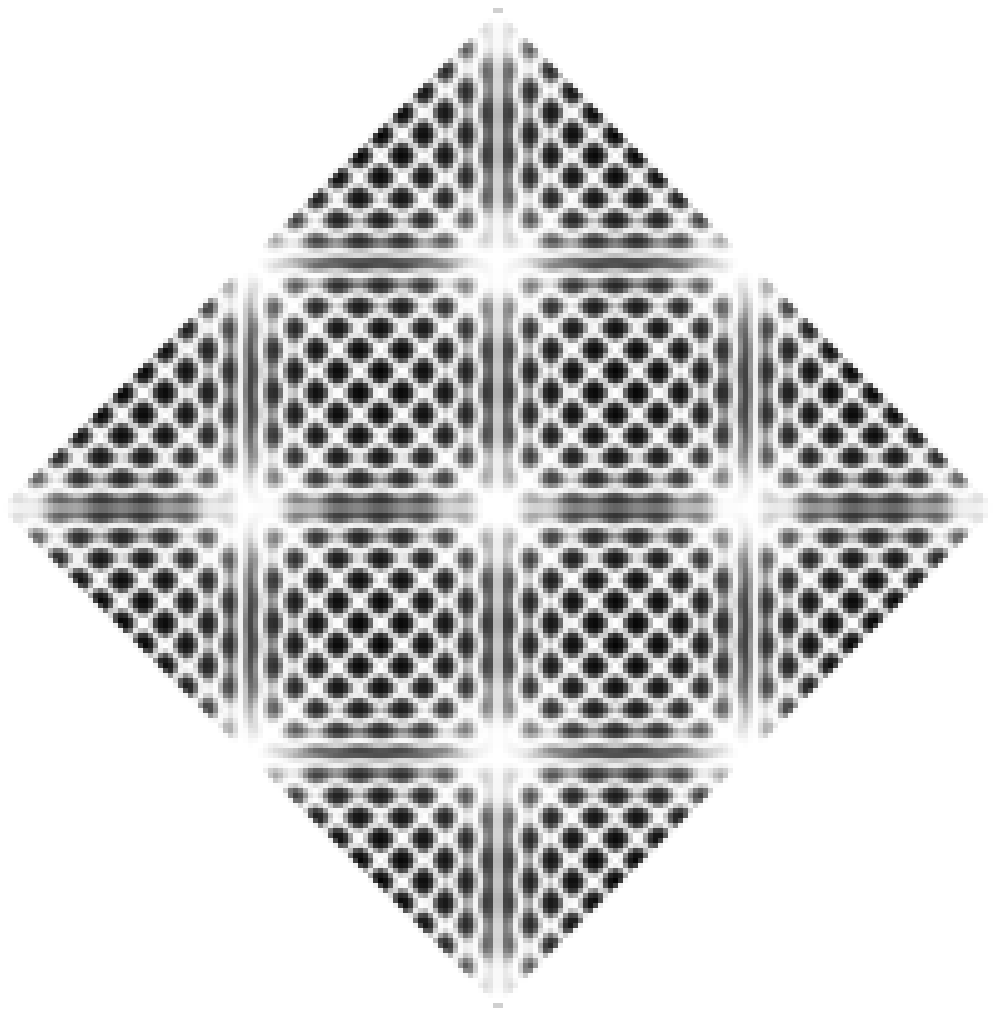}
\end{center}
\begin{center}(b)\end{center}
\end{minipage}\hfill
\begin{minipage}[t!]{.33\linewidth}
\begin{center}
\includegraphics[width=.79\linewidth]{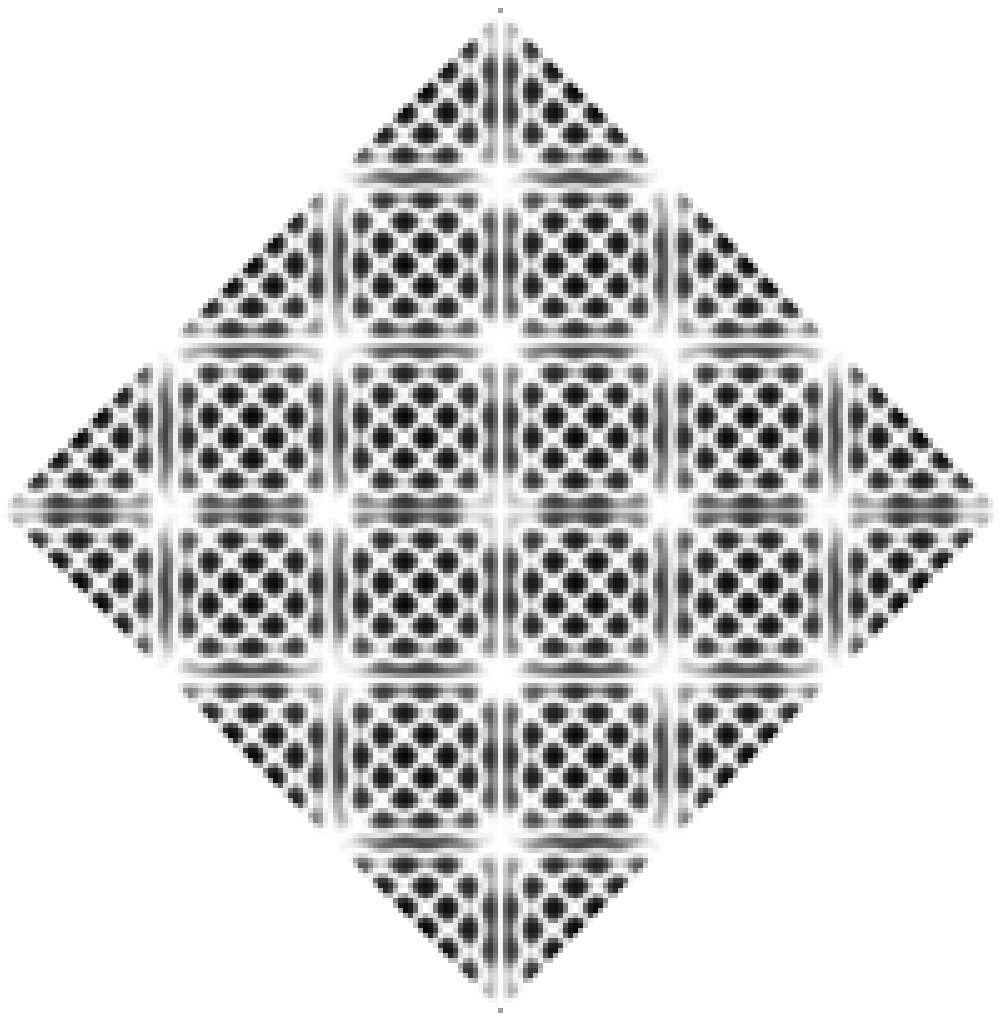}
\end{center}
\begin{center}(c)\end{center}
\end{minipage}
\caption{Squared modulus of wave functions calculated within the
superscar model (\ref{minus_minus}) and corresponding to the
parameters of Fig.~\ref{psi}.} \label{psi_theor}
\end{minipage}
\end{figure*}

These quasi-stationary states are clearly organized in families.
This effect is more pronounced when wave functions corresponding to
each family are calculated. For instance,  wave functions for the
three lowest families with $(-\; -)$ symmetry are presented in
Fig.~\ref{psi}. The other members of these families have similar
patterns. The existence of such families was firstly noted in
\cite{braunhex} for hexagonal dielectric cavities, then further
detailed in \cite{wiersig}.

\begin{figure}[ht!]
  \centering
   \includegraphics[width=0.6\linewidth]{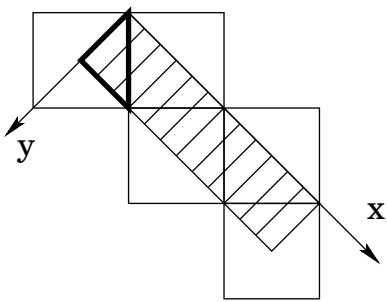}\hfill
    \caption{Unfolding of the diamond periodic orbit. Thick lines indicate the initial triangle.}
    \label{carreorbite}
\end{figure}

 One can argue that the origin of such families is analogous
to the formation of superscar states in pseudo-integrable billiards
discussed in \cite{superscar} and observed experimentally in
microwave experiments in \cite{richter}. In general,  periodic
orbits of polygonal cavities  form continuous families which   can
be considered as propagating inside straight channels obtained by
unfolding classical motion (see Fig.~\ref{carreorbite}). These
channels (hatched area Fig.~\ref{carreorbite}) are restricted by
straight lines passing through cavity corners. In \cite{superscar}
it was demonstrated that strong quantum mechanical diffraction on
these singular corners forces wave functions  in the semiclassical
limit to obey simple boundary conditions  on these (fictitious)
channel boundaries. More precisely it was shown that for billiard
problems $\Psi$ on these boundaries take values of the order of
${\cal O}(1/\sqrt{k})\to 0$ when $k\to \infty$. This result was
obtained by using the exact solution for the scattering on periodic
array of half-planes. No such results are known for dielectric
problems. Nevertheless, it seems natural from semiclassical
considerations that a similar phenomenon should appear for
dielectric polygonal cavities as well.

Within such framework, a superscar state can be constructed
explicitly as follows. After unfolding (see Fig. \ref{carreorbite}),
a periodic orbit channel has the form of a rectangle. Its length
equals the periodic orbit length and its width is determined by the
positions of the closest singular corners. The unfolded superscar
state corresponds to a simple plane wave propagating inside the
rectangle taking into account all phase changes. It cancels at the
fictitious boundaries parallel to the $x$ direction and is periodic
along this direction with a periodicity imposed by the chosen
symmetry class. This procedure sets the wavenumber of the state and
the true wavefunction is obtained by folding back this superscar
state.

Superscar wave functions  with $(-\;-)$ symmetry associated with the
diamond orbit  (see Fig.~\ref{carreorbite}) are expressed as
follows:
\begin{eqnarray}
\Psi_{m,p}^{(-\;-)}(x,y)&=&\sin \left (\kappa_m^{(-)} x\right )\sin
\left (\frac{2\pi}{l}p y \right)\nonumber\\&+& \sin \left
(\kappa_m^{(-)} x'-2\delta\right ) \sin \left (\frac{2\pi}{l}p
y'\right )\; \label{minus_minus}
\end{eqnarray}
where $x'$ and $y'$ are coordinates symmetric with respect to square side. In
coordinates as in Fig.~\ref{carreorbite}
$$x'=y\;,\;\;\; y'=x\;$$
In (\ref{minus_minus})  $m$ and $p$ are integers with
$p=1,2,\ldots,$ and $m\gg 1$. $l=\sqrt{2}a$ is the half of the
diamond periodic orbit length \footnote{For a given symmetry class,
the length entering the
  quantization condition may be a part of the total periodic orbit length.},
$\delta$ is the phase of the reflection coefficient defined by
$r=\exp(-2{\rm i}\delta)$. For simplicity, we ignore slight changes
of the reflection coefficient for different plane waves in the
functions above. So $\delta$ is given by (\ref{phase}) with $\nu=1$
for TM polarization and $\theta=\pi/4$. And $\kappa_m^{(-)}$ is the
momentum defined by
\begin{equation}
\kappa_m^{(-)} l-4\delta=2\pi m\;.
\end{equation}
This construction conducts to the following expression for the real
part of the wavenumbers \footnote{The estimation of the imaginary
parts of these states as well as the field distribution outside the
cavity is beyond the scope of this paper and will be discussed
elsewhere. }
\begin{eqnarray}
n_{eff}l{\rm Re}(k_{m,p} )&=&
2\pi\sqrt{(m+\frac{2}{\pi}\delta)^2+p^2}\nonumber\\
&=&2\pi(m+\frac{2}{\pi}\delta) +{\cal O}(\frac{1}{m})\;.
\label{Emm}
\end{eqnarray}
To check the accuracy of the above formulae we plot in
Fig.~\ref{psi_theor}  scar wave functions (\ref{minus_minus})  with
the same parameters as those in Fig.~\ref{psi}. The latter were
computed numerically by direct solving the Helmholtz equations
(\ref{Eqs}) but the former looks very similar which supports the
validity of the superscar model.

The real part of the wavenumbers is tested too. In Fig.~\ref{total}
(b) the lowest loss states (with the smallest modulus of the
imaginary part) with $(-\;-)$ symmetry are presented over a larger
interval than in Fig.~\ref{total} (a). The real parts of these
states are compared to superscar predictions (\ref{Emm}) with $p=1$,
leading to a good agreement. To detect small deviations from the
theoretical formula, we plot in the inset of Fig.~\ref{total} (b)
the difference between a quantity inferred from numerical
simulations
 and its superscar prediction from (\ref{Emm}).
\begin{equation}
\delta E=\left (\frac{n l}{2\pi} {\rm Re }k\right )^2-\left ((m+\frac{2}{\pi}\delta
  )^2+p^2\right )\;.
\label{deltaE}
\end{equation}
From this curve it follows that this difference tends to zero with
$m$ increasing, thus confirming the existence of the term
proportional to $p^2$. By fitting this difference with the simplest
expression
\begin{equation}
\delta E=\frac{c}{m}
\label{correction}
\end{equation}
we find that $c\simeq -6.9$. By subtracting this correction term
from the difference (\ref{deltaE}), one gets the curve indicated
with filled circles in inset of  Fig.~\ref{total} (b). The result is
one order of magnitude smaller than the difference itself.

All these calculations confirm that the real parts of resonance
wavenumbers for square dielectric cavities are well reproduced in
the semiclassical limit by the above superscar formula (\ref{Emm})
and our experimental results can be considered as an implicit
experimental confirmation of this statement.

\section{Pentagonal micro-cavity}\label{pentagonal_cavity}

\begin{figure}
\begin{minipage}[t!]{0.5\linewidth}
\includegraphics[ width=0.7\linewidth]{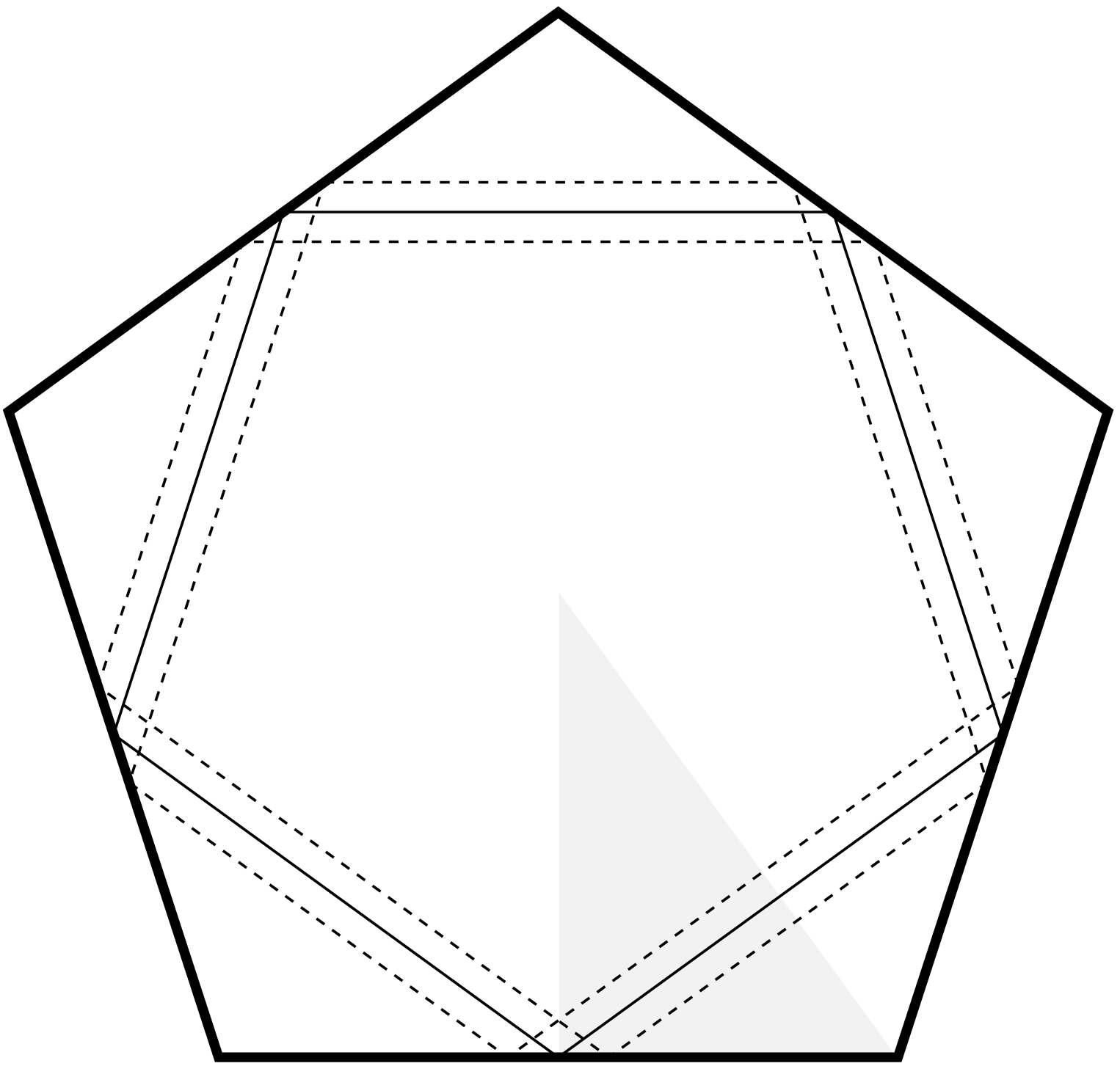}
\begin{center}(a)\end{center}
\end{minipage}\hfill
\begin{minipage}[t!]{0.5\linewidth}
\includegraphics[width=0.7\linewidth]{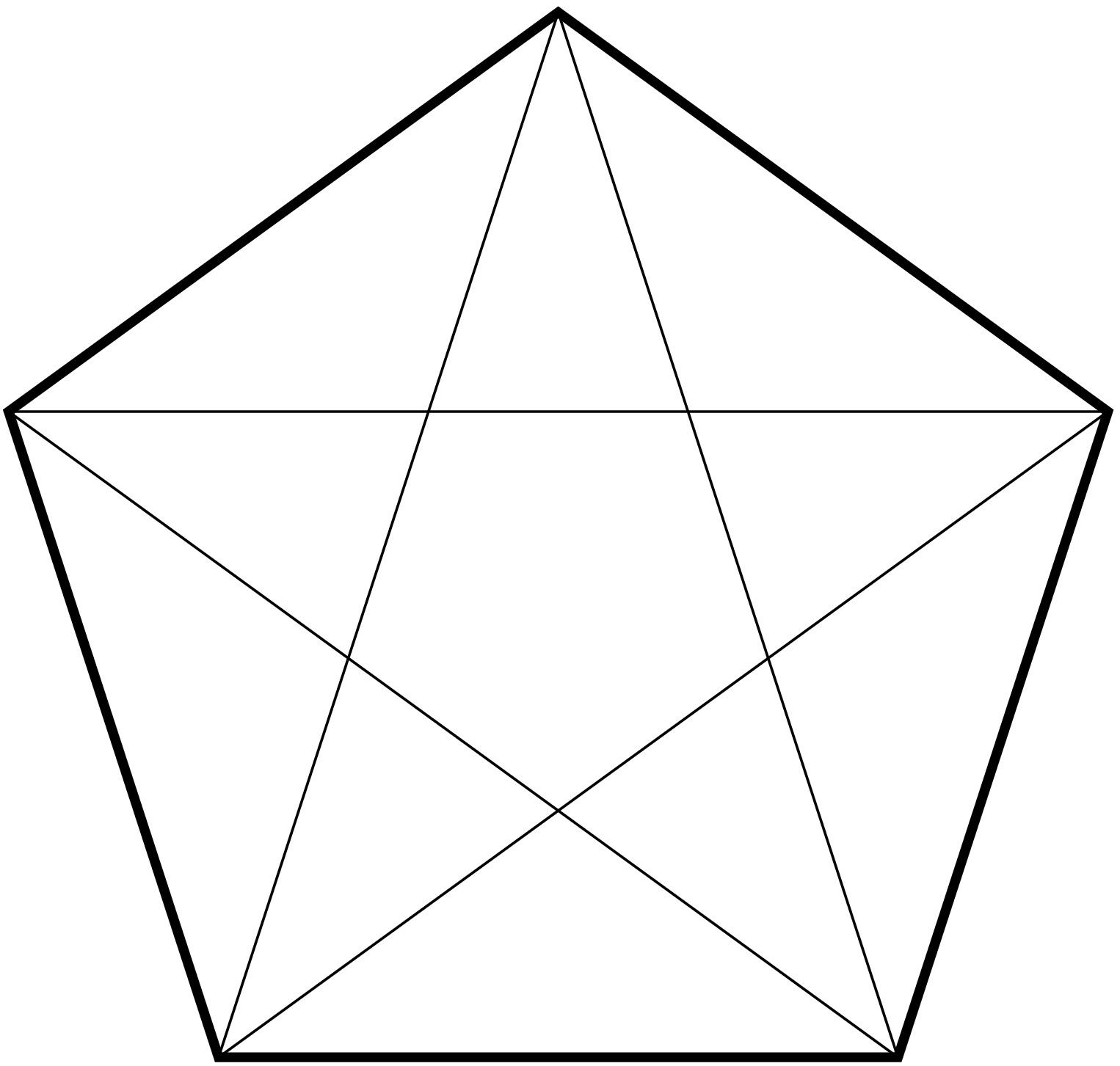}
\begin{center}(b)\end{center}
\end{minipage}
\caption{Simplest whispering gallery periodic orbit family for a  pentagonal
    cavity. (a) Solid line indicates the inscribed pentagon which is an
    isolated periodic orbit. A periodic orbit in its vicinity is plotted
    with
    dashed line. It belongs to the family of the five-pointed star periodic
    orbit.
        The fundamental domain is indicated in grey.
    (b) Boundary of the family of the five-pointed star periodic orbit.}
    \label{pentaorbites}
\end{figure}

The trace formula and  superscar model arguments can be generalized
to all polygonal cavities. The pentagonal resonator provides a new
interesting test. In fact, due to the odd number of sides, the
inscribed pentagonal orbit (indicated by solid line in
Fig.~\ref{pentaorbites} (a)) is isolated. The shortest confined
periodic orbit family is twice longer. It is represented with a
dashed line in Fig.~\ref{pentaorbites} (a) and can be mapped onto
the five-pointed star orbit drawn in Fig.~\ref{pentaorbites} (b) by
continuous deformation. In this Section we compare the predictions
of the superscar model for this periodic orbit family with numerical
simulations and experiments.

\begin{figure}
\begin{minipage}[t!]{\linewidth}
\includegraphics[width=.8\linewidth]{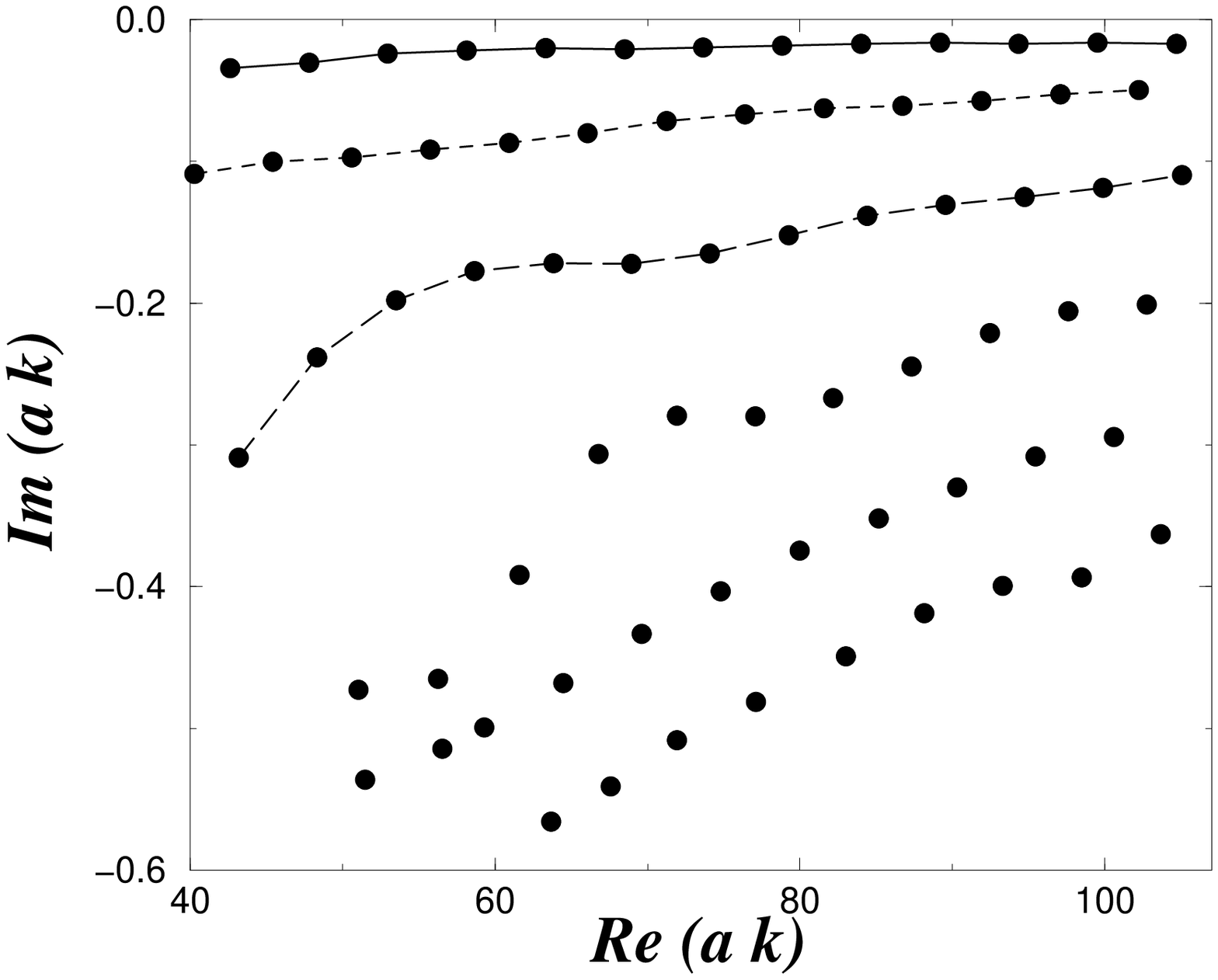}
\begin{center}(a)\end{center}
\end{minipage}
\vspace{1em}
\begin{minipage}[t!]{\linewidth}
\includegraphics[ angle=-90, width=.8\linewidth]{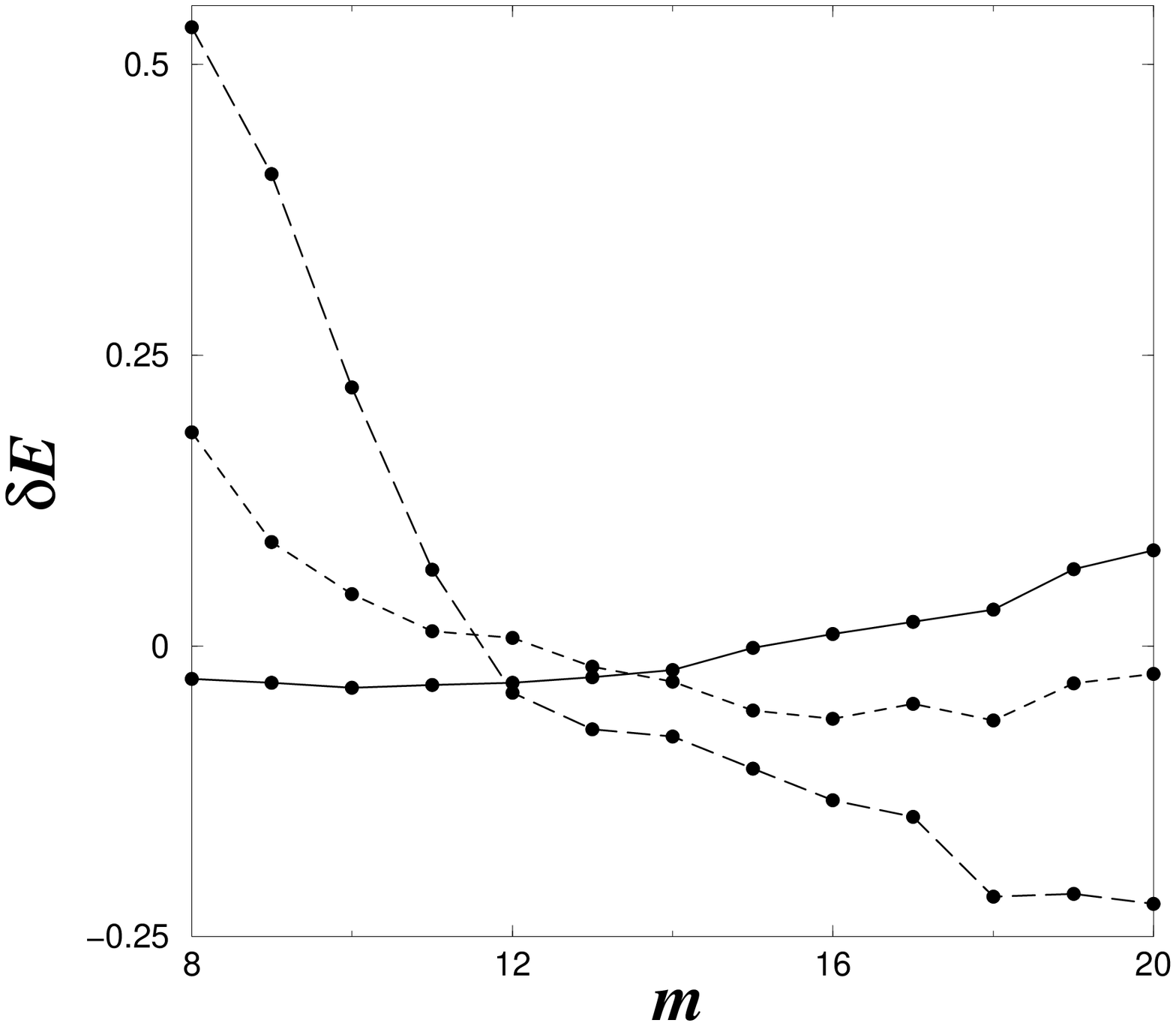}
\begin{center}(b)\end{center}
\end{minipage}
\caption{(a)
 Wave numbers for  a pentagonal cavity. $a$ is the side length of the cavity.
 The three most confined families are indicated by solid, dashed
 and long-dashed lines. (b) The difference (\ref{difference}) between the real part of
 quasi-energies  and superscar expression (\ref{penta_scar}) for the
 three indicated families in (a). }
\label{penta}
\end{figure}

Due to the $C_{5v}$ symmetry, pentagonal cavities sustain 10
symmetry classes corresponding to the rotations by $2\pi/5$ and the
inversion with respect to one of the symmetry axis. We have studied
numerically  one symmetry class in which wave functions obey the
Dirichlet boundary conditions along two sides of a right triangle
with angle $\pi/5$ (see Fig.~\ref{pentaorbites} (a) in grey).  The
results of these computations are presented in Fig.~\ref{penta}.

\begin{figure*}
\begin{minipage}{1.\linewidth}
\begin{minipage}[t!]{.33\linewidth}
\includegraphics[width=.99\linewidth, angle=-90]{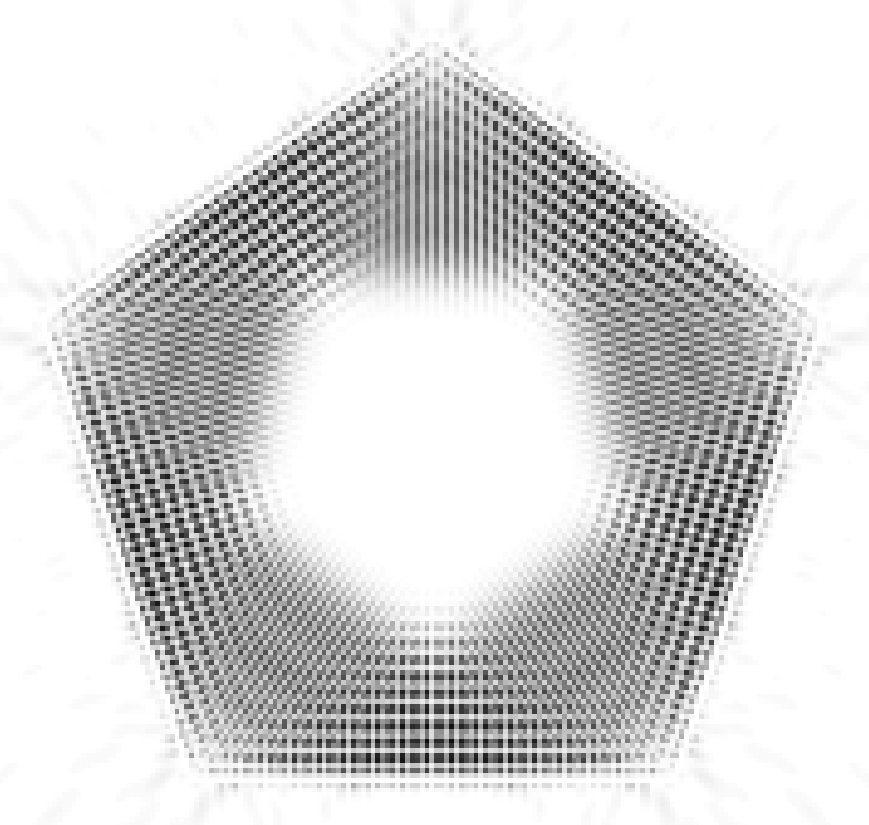}
\begin{center}(a)\end{center}
\end{minipage}\hfill
\begin{minipage}[t!]{.33\linewidth}
\includegraphics[width=.99\linewidth, angle=-90]{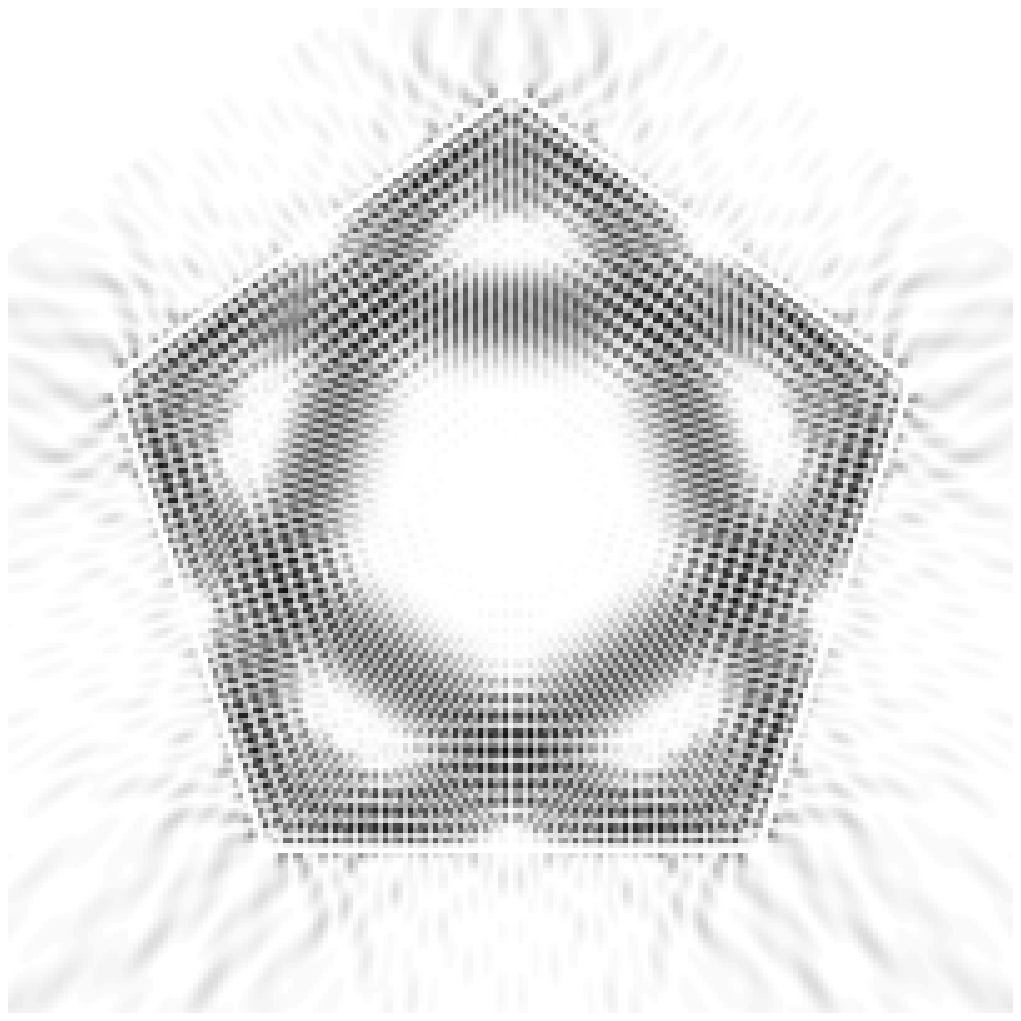}
\begin{center}(b)\end{center}
\end{minipage}\hfill
\begin{minipage}[t!]{.33\linewidth}
\includegraphics[width=.99\linewidth, angle=-90]{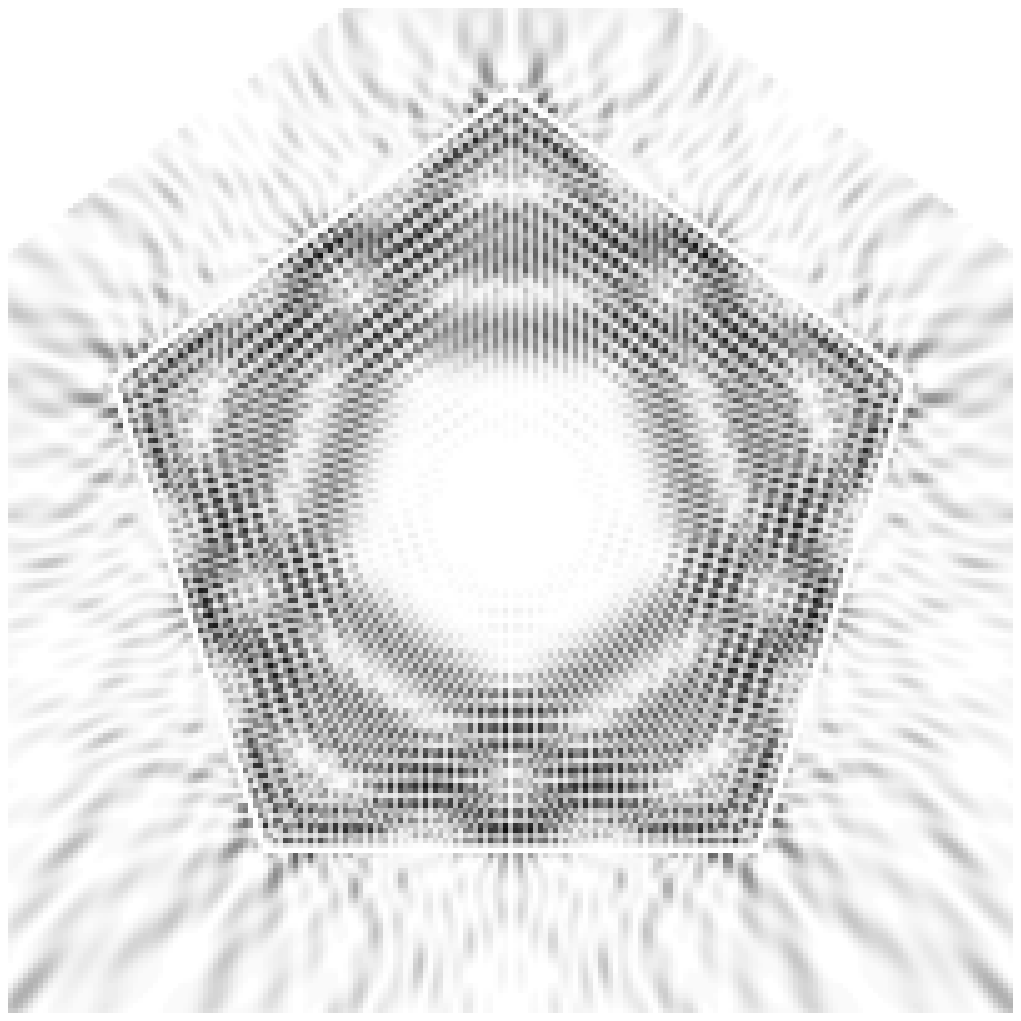}
\begin{center}(c)\end{center}
\end{minipage}
\caption{Squared modulus of wave functions for pentagonal cavity
with $(-\;-)$ symmetry calculated with numerical simulations. (a)
  $ak=104.7-0.017~{\rm i}$, (b) $ak=102.2-0.05~{\rm i}$,
  (c) $ak=105.0-0.12~{\rm  i}$.   }
\label{psi_penta}
\begin{minipage}[t!]{.33\linewidth}
\begin{center}
\includegraphics[width=.79\linewidth, angle=90]{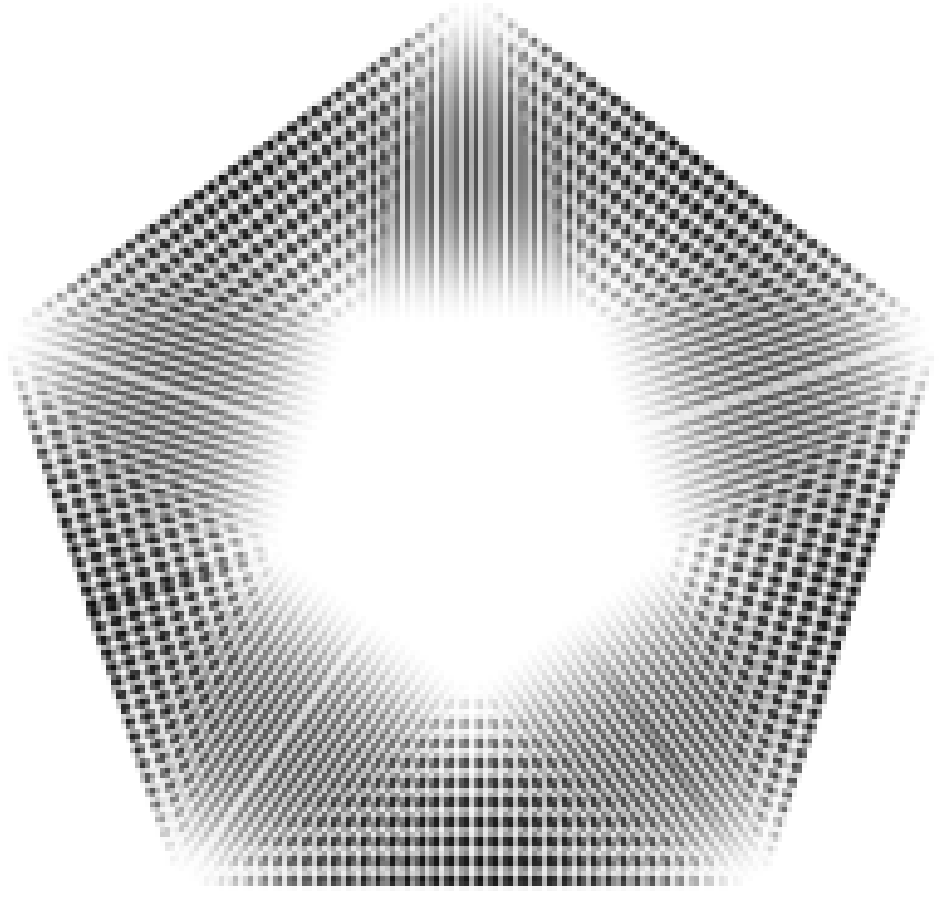}
\end{center}
\begin{center}(a)\end{center}
\end{minipage}\hfill
\begin{minipage}[t!]{.33\linewidth}
\begin{center}
\includegraphics[width=.79\linewidth, angle=90]{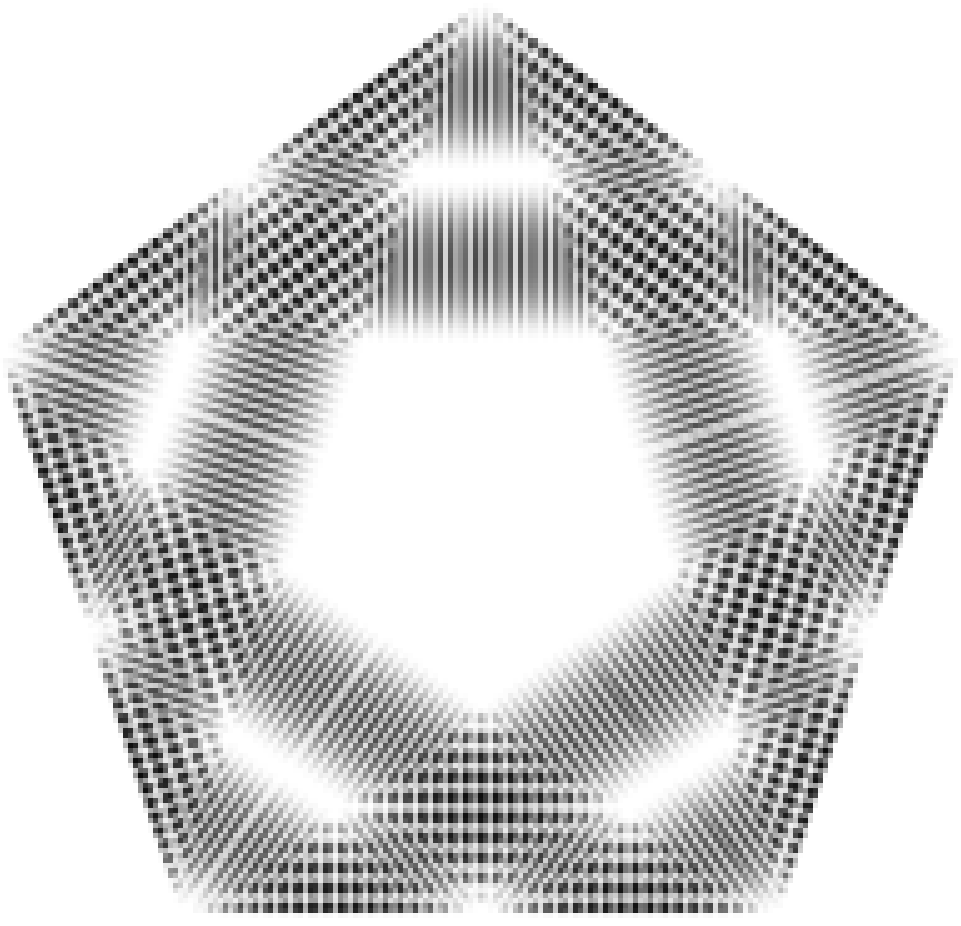}
\end{center}
\begin{center}(b)\end{center}
\end{minipage}\hfill
\begin{minipage}[t!]{.33\linewidth}
\begin{center}
\includegraphics[width=.79\linewidth, angle=90]{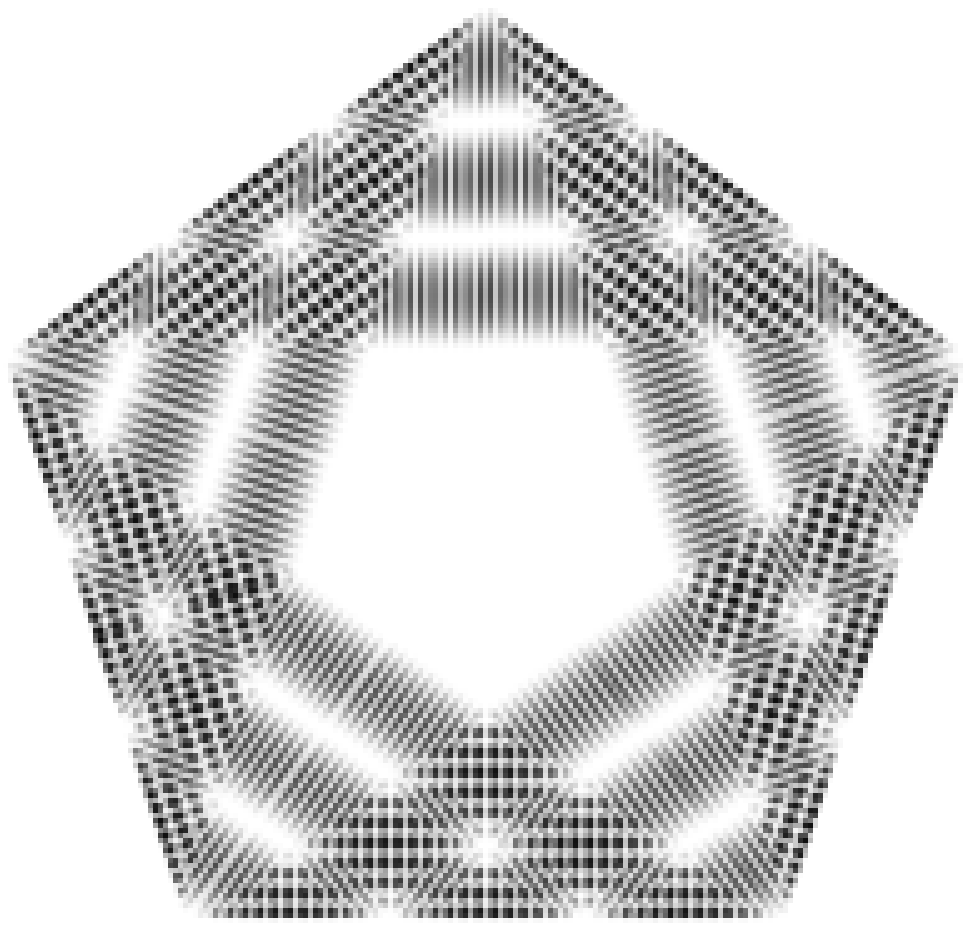}
\end{center}
\begin{center}(c)\end{center}
\end{minipage}
\caption{Squared modulus of wave functions calculated within the
superscar model and corresponding to the parameters of
Fig.~\ref{psi_penta}.} \label{psi_penta_theor}
\end{minipage}
\end{figure*}

As for the square cavity, lowest loss states are organized in
families. The wave functions of the three lowest loss families are
plotted in Fig.~\ref{psi_penta} and their superscar structure is
obvious.

\begin{figure}[ht!]
  \centering
   \includegraphics[width=1.\linewidth]{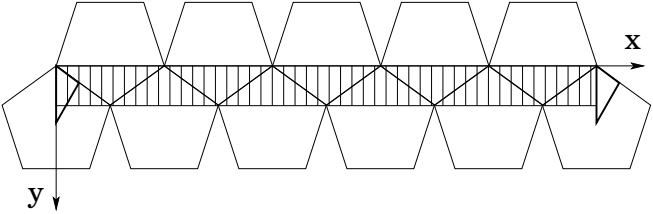}
    \caption{Unfolding of the five-star periodic orbit for a pentagonal
      cavity. Thick lines indicate the initial triangle.}
    \label{pentaorbite}
\end{figure}

The computation of pure superscar states can be performed as in the
previous Section. The five-pointed star periodic orbit channel is
shown in Fig.~\ref{pentaorbite}. In this case boundary conditions
along horizontal boundaries of periodic orbit channel are not known.
By analogy with superscar formation in polygonal billiards
\cite{superscar}, we impose that wave functions tend to zero along
these boundaries when $k\to\infty$.

Therefore, a superscar wave function propagating
inside this channel takes the form
\begin{equation}
\Psi_{{\rm scar}}(x,y)=\exp({\rm i}\kappa x)\sin(\frac{\pi}{w}py)
\Theta(y)\Theta(w-y)\;, \label{scar_general}
\end{equation}
where $w$ is the width of the channel (for the five-pointed star
orbit $w=a\sin(\pi/5)$ where $a$ is the length of the pentagon
side). $\Theta(x)$ is the Heavyside function introduced here to
stress that superscar functions are zero (or small) outside the
periodic orbit channel.

The quantized values of the longitudinal momentum, $\kappa$,  are
obtained by imposing that the function (\ref{scar_general}) is
periodic along the channel when all phases due to the reflection
with the cavity boundaries are taken into account
\begin{equation}
\kappa L=2\pi\left ( M+\frac{10}{\pi}\delta \right )\;.
\label{penta_M}
\end{equation}
Here $M$ is an integer and $L$ is the total periodic orbit length.
For the five-pointed star  orbit (see Fig.~\ref{pentaorbites})
\begin{equation}
L=10 a \cos (\frac{\pi}{5})\;,
\end{equation}
and $\delta$ is the phase of the reflection coefficient given by
(\ref{phase}) with $\nu=1$ (for TM polarization) and
$\theta=3\pi/10$. For these states the real part of the wavenumber
is the following
\begin{equation}
n L {\rm  Re } k=2\pi\left ( M+\frac{10}{\pi}\delta \right )+{\cal
O}(\frac{1}{M})\;. \label{penta_scar}
\end{equation}
Wave function inside the cavity are obtained by folding back the
scar function (\ref{scar_general}) and choosing the correct
representative of the chosen symmetry class. When Dirichlet boundary
conditions are imposed along two sides of a right triangle passing
through the center of the pentagon (see Fig.~\ref{pentaorbite}), $M$
must be written as $M=5(2m)$ if $p$ is odd and $M=5(2m-1)$ if $p$ is
even. Then the wave function inside the triangle is the sum of two
terms
\begin{eqnarray}
&&\Psi_{m,p}(x,y)=\sin(\kappa_m x)\sin(\frac{\pi}{w}py)
\Theta(y)\Theta(w-y)+\nonumber\\
&+&\sin(\kappa_m x'-2\delta)\sin(\frac{\pi}{w}py')
\Theta(y')\Theta(w-y') \label{scar_symmetry}
\end{eqnarray}
where the longitudinal momentum is
\begin{equation}\label{longitudinalpenta}
\kappa_m\frac{L}{10}=2\pi(m+\frac{1}{\pi}\delta-\xi)
\end{equation}
with $\xi=0$ for odd $p$ and $\xi=1/2$ for even $p$. $x'$ and $y'$
in (\ref{scar_symmetry}) are coordinates of the point symmetric of
$(x,y)$ with respect to the inversion on the edge of the pentagon.
In the coordinate system when the pentagon edge passes through the
origin (as in Fig.~\ref{pentaorbite})
$$x'=x\cos 2\phi +y\sin 2\phi\;,\;\;\; y'=x\sin 2\phi -y\cos 2\phi$$
and $\phi=\pi/5$ is the inclination angle of the pentagon side with
respect to the abscissa axis. Wavefunctions obtained with this
construction  are presented in Fig.~\ref{psi_penta_theor}. They
correspond to the first, second, and third perpendicular excitations
of the five-star periodic orbit family ($p$~= 1, 2, and 3).

To check the agreement between numerically computed real parts of
the wavenumbers and the superscar prediction (\ref{penta_scar}) and
(\ref{longitudinalpenta}), we plot in Fig.~\ref{penta} (b) the
following difference
\begin{eqnarray}
\delta {\rm }E&=&\left (\frac{n\cos\phi}{2\pi} {\rm  Re}ka_{num}\right )^2-\nonumber\\
&-&\left [\left (m+\frac{1}{\pi}\delta -\xi \right )^2+\left
    (\frac{1}{2\tan\phi}\right )^2\zeta\right ] \;.
\label{difference}
\end{eqnarray}
For pure scar states $\zeta=p^2$. As our numerical simulations have
not reached the semiclassical limit (see scales in Figs.~\ref{total}
and \ref{penta}), we found it convenient to fit numerically the
$\zeta$ constant.  The best fit gives $\zeta \approx$ 0.44, 2.33,
and 5.51 for the three most confined families (for pure scar
functions this constant is 1, 4, 9 respectively). The agreement is
quite good with a relative accuracy of the order of $10^{-4}$  (see
Fig.~\ref{penta} (b)). Irrespective of precise value of $\zeta$ the
total optical length, $nL$, is given by (\ref{penta_scar}) and leads
to an experimental prediction twice longer than the optical length
of the inscribed pentagon, which is an isolated periodic orbit and
thus can not base superscar wavefunctions.\\

\begin{figure}[ht!]
  \centering
   \includegraphics[width=1\linewidth]{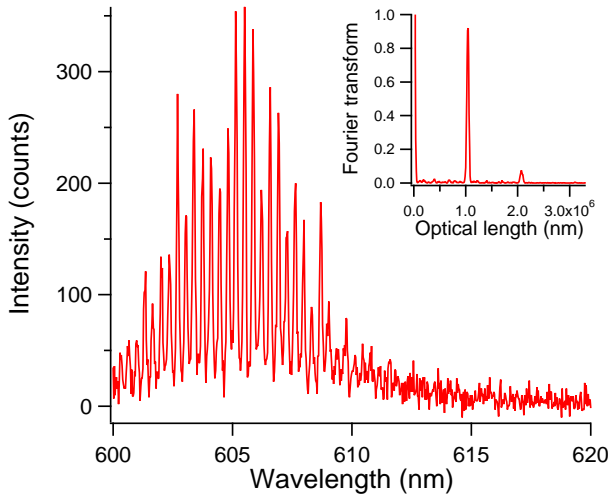}\hfill
    \caption{Experimental spectrum of a pentagonal micro-laser of
$80~\mu m$ side length. Inset: Normalized Fourier transform of the
spectrum expressed as intensity versus wavenumber.}
    \label{pentamanip}
\end{figure}

Comparison with experiments confirms the superscar nature of the
most confined states for pentagonal resonators. In fact, the
spectrum and its Fourier transform in Fig.~\ref{pentamanip}
correspond to a pentagonal micro-laser with side $a=80~\mu m$, and
show a periodic orbit with optical length $1040~\pm~30~\mu m$ to be
compared with the five-star optical length
$n_{full}10a\cos(\pi/5)=1061~\mu m$. The agreement is better than
2\%.\\
This result is reproducible for cavities with the same size. Other
sizes have been tested as well. For smaller cavities, the
five-pointed star orbit is not identifiable due to lack of gain,
whereas for bigger ones it is visible but mixed with non confined
periodic orbits. This effect, not specific to pentagons, can be
assigned to the contribution of different periodic orbit families
which become important when the lasing gain exceeds the refractive
losses. We will describe this phenomenon in a future publication
\cite{nouscompetitionorbites}.

The good agreement of numerical simulations and experiments
with superscar predictions gives an additional credit to the validity of
this approach even for non-trivial configurations.

\section{Micro-disks}\label{circular_cavity}

Micro-disk cavities are the simplest and most widely used
micro-resonators. In the context of this work, they are of interest
because of the coexistence of several periodic orbit families with
close lengths. For low index cavities ($n\sim 1.5$) each regular
polygon trajectory with more than four sides is confined by total
internal reflection.

In the two-dimensional approximation passive circular cavities are integrable
and the spectrum of quasi-stationary states can be computed from an
explicit quantization condition
\begin{equation}\label{eqdisque}
n\frac{J_m^{\prime}(n kR)}{J_m(n kR)}=\nu
\frac{H_m^{(1)\prime}(kR)}{H_m^{(1)}(kR)}\;.
\end{equation}
Here $R$ is the radius of the disk, $n$ the refractive index of the
cavity, and $\nu=1$ (resp. $\nu=n^2$)  for the TM (resp. TE)
polarization. For each angular quantum number $m$, an infinite
sequence of solutions, $k_{m,q}$, is deduced from (\ref{eqdisque}).
They are labeled by the $q$ radial quantum number.

\begin{figure}
\begin{minipage}[t!]{0.5\linewidth}
\includegraphics[width=0.6\linewidth]{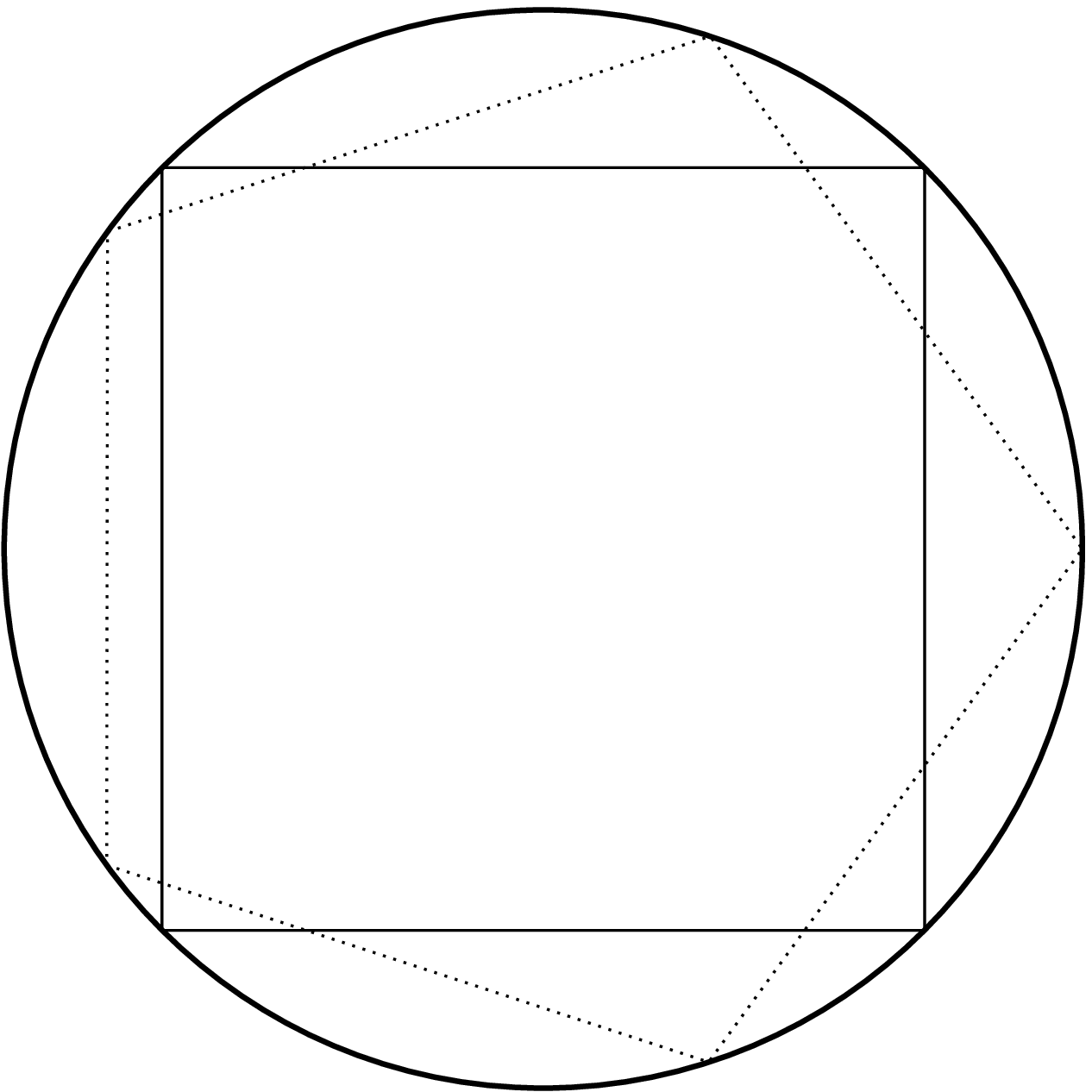}
\begin{center}(a)\end{center}
\end{minipage}\hfill
\begin{minipage}[t!]{0.5\linewidth}
\includegraphics[width=0.6\linewidth]{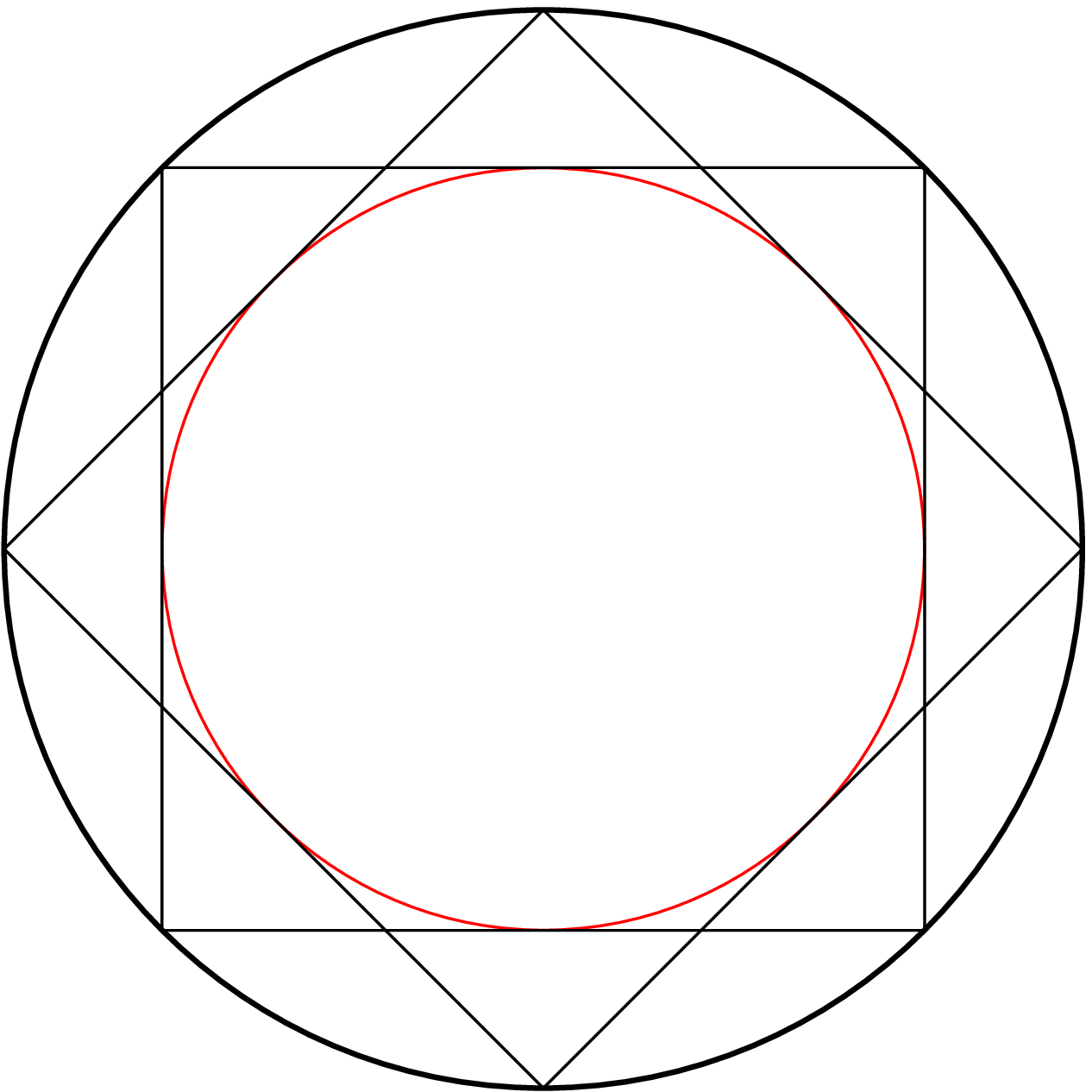}
\begin{center}(b)\end{center}
\end{minipage}
\caption{(a) Two examples of periodic orbits: the square and the
    pentagon. (b) Two representations of the square periodic orbit and
    the caustic of this family in red.}
    \label{disqueorbites}
\end{figure}

\begin{figure}[ht!]
  \centering
   \includegraphics[width=0.7\linewidth]{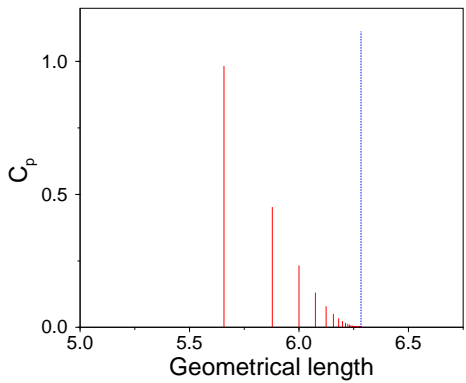}\hfill
    \caption{Vertical red sticks: $c_p$ coefficient for polygons
confined by total internal reflection (square, pentagon, hexagon,
etc...).
 The dotted blue line indicates the position of the perimeter.}
    \label{disquetrace}
\end{figure}

For large $|k|$ the $k_{m,l}$ wavenumbers are obtained from a
semiclassical expression (see e.g. \cite{nockelthese}) and the
density of quasi-stationary states  (\ref{quasi_density}) can be
proved to be rewritten  as a sum over periodic orbit families. The
derivation of this trace formula assumes only the semi-classical
approximation ($|k|R\gg 1$) and can be done in a way similar to that
of the billiard case (see e.g. \cite{brack}), leading to an
expression closed to (\ref{trace_k})
\begin{equation}\label{traceouverte}
d(k)~\propto \sum_{p} \frac{A_p}{\sqrt{L_p}} ~|r_p|^{N_p}~
\cos(nL_p k-N_p~(2\delta_p+\frac{\pi}{2})+\frac{\pi}{4})\;.
\end{equation}
Here the $p$ index specifies a periodic orbit family. This formula
depends on periodic orbit parameters: the number of bounces on the
boundary, $N_p$, the incident angle on the boundary, $\chi_p$, the
length, $L_p=2N_p R\cos(\chi_p)$, and the area covered by periodic
orbit family, $A_p=\pi R^2\cos^2(\chi_p)$, which is the area
included between the caustic and the boundary (see Fig.
\ref{disqueorbites} (b)). $2\delta_p$ is the phase of the reflection
coefficient at each bounce on the boundary (see Eq. (\ref{phase}))
and
$|r_p|$ is its modulus.\\
For  orbits confined by total internal reflection $\delta_p$ does
not depend on $kR$ in the semi-classical limit, and $r_p$ is
exponentially close to 1 \cite{nockelthese,martina}. From
(\ref{traceouverte}) it follows that  each periodic orbit is singled
out by a  weighing coefficient $c_p=\frac{A_p}{\sqrt{L_p}}
~|r_p|^{N_p}$. Considering the experimental values $|k|R\sim 1000$,
$|r_p|$ can be approximated to unity with a good accuracy for
confined periodic orbits, and thus $c_p=\frac{A_p}{\sqrt{L_p}}$
depends only on geometrical quantities. Fig.~\ref{disquetrace} shows
the evolution of $c_p$ for polygons when the number of sides is
increasing. As the critical angle is close to $45^{\circ}$, the
diameter and triangle periodic orbits are not confined and the
dominating contribution comes from the square periodic orbit. So we
can reasonably conclude that the spectrum (\ref{quasi_density}) of a
passive two-dimensional
micro-disk  is dominated by the square periodic orbit.\\

\begin{figure}[ht!]
\begin{minipage}[ht!]{1.\linewidth}
\includegraphics[width=0.8\linewidth]{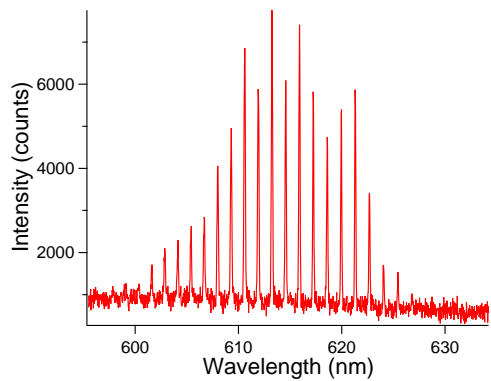}
\begin{center}(a)\end{center}
\end{minipage}\hfill
\begin{minipage}[ht!]{1.\linewidth}
\includegraphics[width=0.8\linewidth]{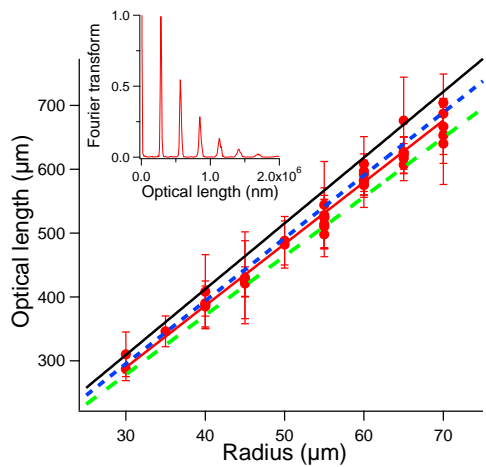}
\begin{center}(b)\end{center}
\end{minipage}
\caption{(a) Experimental spectrum of a micro-disk of $30~\mu m$
radius.
    (b) Optical length versus radius. The experiments (red points) are
       linearly fitted by the solid red line. The other lines
       correspond to theoretical predictions without any adjusted
       parameters: the dashed green line to the square, the dotted
       blue line to the hexagon, and the solid black line to the
       perimeter.
       Inset: Normalized Fourier transform of the spectrum in (a) expressed as
       intensity versus wavenumber.}
    \label{disqueresultat}
\end{figure}

The experimental method described in the previous Sections has been
applied to disk-shaped micro-cavities. A typical experimental
spectrum is shown on Fig.~\ref{disqueresultat} (a). The first peak
of its Fourier transform (see Fig. \ref{disqueresultat} (b) inset)
has a finite width coming  from the experimental conditions
(discretization, noise, etc...) and the contributions of several
periodic orbits. This width is represented as error bars on graph
\ref{disqueresultat} (b). The continuous red line fitting the
experimental data is surrounded by the dashed green line and the
dotted blue line corresponding to the optical length of the square
and hexagon respectively, calculated with $n_{full}=1.64$ as in the
previous Sections. The perimeter (continuous black line) overlaps
with a large part of the error bars which evidences its contribution
to the spectrum, but it is not close to experimental data.

These experimental results seem in good agreement with the above
theoretical predictions. But actually these resonances, usually
called whispering gallery modes, are living close to the boundary.
Thus both roughness and three-dimensional effects must be taken into
account. At this stage it is difficult to evaluate and to measure
correctly such contributions for each periodic orbit. For
micro-disks with a small thickness (about 0.4 $\mu$m) and designed
with lower roughness, the results are more or less similar to those
presented in Fig.~\ref{disqueresultat} (b).

\section{Conclusion}

We demonstrate experimentally that the length of the dominant
periodic orbit can be recovered from the spectra of micro-lasers
with simple shapes. Taking into account different dispersion
corrections to the effective refractive index,  a good agreement
with theoretical predictions has been evidenced first for the
Fabry-Perot resonator. Then we have tested polygonal cavities both
with experiments and numerical simulations, and a good agreement for
the real parts of wavenumbers has been obtained even for the non
trivial configuration of the pentagonal cavity.\\
The observed dominance of confined short-period orbits is, in
general, a consequence of the trace formula and the formation of
long-lived states in polygonal cavities is related to strong
diffraction on cavity corners.\\
Finally, the study of micro-disks highlights the case of several
orbits and the influence of roughness and three-dimensional effect.

Our study opens the way to a systematic exploration of spectral
properties by varying the shape of the boundary. In increasing the
experimental precision even tiny details of trace formulae will be
accessible. The improvement of the etching quality will suppress the
leakage due to surface roughness and lead to a measure of the
diffractive mode losses which should depend on symmetry classes.
From the point of view of technology, it will allow a better
prediction of the resonator design depending on the applications.
From a more fundamental physics viewpoint, it may contribute to a
better understanding of open dielectric billiards.

\section{Acknowledgments}

The authors are grateful to S. Brasselet, R. Hierle, J. Lautru, C.
T. Nguyen, and J.-J. Vachon for experimental and technological
support and to C.-M. Kim, O. Bohigas, N. Sandeau, J. Szeftel, and E.
Richalot for fruitful discussions.


\end{document}